\documentclass[iop,revtex4]{emulateapj}
\usepackage{amsmath}
\usepackage{color}

\providecommand{\HII}{H~{\footnotesize II}}			
\providecommand{\OIII}{[O~{\footnotesize III}]}			
\providecommand{\OII}{[O~{\footnotesize II}]}			
\providecommand{\SII}{[S~{\footnotesize II}]}			
\providecommand{\NII}{[N~{\footnotesize II}]}			
\providecommand{\HA}{H$\alpha$}					
\providecommand{\HB}{H$\beta$}					
\providecommand{\R}{$R_{23}$}					
\providecommand{\Te}{$T_{e}$}					
\providecommand{\Ne}{$n_{e}$}					
\providecommand{\abun}{12+$\log$(O/H)}				
\providecommand{\LZ}{$L$--$Z$}					
\providecommand{\MZ}{$M_{*}$--$Z$}				
\providecommand{\MZSFR}{$M_{*}$--$Z$--SFR}		
\providecommand{\MSFR}{$M_{*}$--SFR}				

\slugcomment{3 January 2018 -- Accepted for Publication in AJ} 

\shorttitle{KISS Metallicity Relations}
\shortauthors{Hirschauer et al.}

\begin{document}


\title{Metal Abundances of KISS Galaxies.\ VI.\ New Metallicity Relations for the KISS Sample of Star-Forming Galaxies}






\author{Alec S.\ Hirschauer\altaffilmark{1,2}, John J.\ Salzer\altaffilmark{1}, Steven Janowiecki\altaffilmark{3}, \& Gary A.\ Wegner\altaffilmark{4}}
\altaffiltext{1}{Department of Astronomy, Indiana University, 727 East Third Street, Bloomington, IN 47405.  {\it e--mail:}  ash@astro.indiana.edu, slaz@astro.indiana.edu}
\altaffiltext{2}{Space Telescope Science Institute, 3700 San Martin Drive, Baltimore, MD 21218.  {\it e--mail:}  ahirschauer@stsci.edu}
\altaffiltext{3}{International Centre for Radio Astronomy Research (ICRAR), University of Western Australia, 35 Stirling Highway, Crawley, WA 6009, Australia.  {\it e--mail:} steven.janowiecki@uwa.edu.au}
\altaffiltext{4}{Department of Physics and Astronomy, Dartmouth College, 6127 Wilder Laboratory, Hanover, NH 03755.  {\it e--mail:} gary.a.wegner@dartmouth.edu}


\begin{abstract}
We present updated metallicity relations for the spectral database of star-forming galaxies (SFGs) found in the KPNO International Spectroscopic Survey (KISS).
New spectral observations of emission-line galaxies (ELGs) obtained from a variety of telescope facilities provide oxygen abundance information.
A nearly four-fold increase in the number of KISS objects with robust metallicities relative to our previous analysis provides for an empirical abundance calibration to compute self-consistent metallicity estimates for all SFGs in the sample with adequate spectral data.
In addition, a sophisticated spectral energy distribution (SED) fitting routine has provided robust calculations of stellar mass.
With these new and/or improved galaxy characteristics, we have developed luminosity-metallicity (\LZ) relations, mass-metallicity (\MZ) relations, and the so-called Fundamental Metallicity Relation (FMR) for over 1,450 galaxies from the KISS sample.
This KISS \MZ\ relation is presented for the first time and demonstrates markedly lower scatter than the KISS \LZ\ relation.
We find that our relations agree reasonably well with previous publications, modulo modest offsets due to differences in the SEL metallicity calibrations used.
We illustrate an important bias present in previous \LZ\ and \MZ\ studies involving direct-method (\Te) abundances that may result in systematically lower slopes in these relations.
Our KISS FMR shows consistency with those found in the literature, albeit with a larger scatter.
This is likely a consequence of the KISS sample being biased toward galaxies with high levels of activity.
\end{abstract}



\section{Introduction} 

\indent The ongoing synthesis of heavy elements and subsequent enrichment of the interstellar medium (ISM) by successive generations of star formation drives the chemical evolution of galaxies populating our universe.
Gas-phase metal abundances of star-forming galaxies (SFGs) can be derived with good precision via nebular spectroscopy.
Oxygen abundance is conventionally used as a proxy for global metallicity, as it is the most prevalent heavy element produced within massive stars.
Furthermore, nebular spectra provide strong emission lines from both primary ionization states (O$^{+}$ and O$^{++}$) in the optical region of the spectrum, allowing for abundance determination without the need for ionization correction factors (ICFs).
\\
\indent Metallicity has long been known to correlate with the overall stellar content of a given system, measured by the integrated luminosity or through an estimate of the total stellar mass, but is additionally sensitive to galactic processes such as infall, outflow, and feedback.
These latter factors are an added complication that necessitate further observational study in order to better constrain models of chemical evolution.
\citet{bib:Lequeux1979} first showed the relationship between stellar mass and metallicity for a local sample of dwarf galaxies, while the correlation between galaxy luminosity and metallicity was first demonstrated by \citet{bib:Rubin1984}.
\citet{bib:Tremonti2004} expanded these correlations using a large number ($\sim$53,400) of objects taken from the Sloan Digital Sky Survey (SDSS; \citealp{bib:York2000, bib:Abazajian2004}).
Many studies have focused on refining these relations, utilizing techniques which include variations in sample selection, metallicity calibration, and theoretical models for comparison (e.g., \citealp{bib:Skillman1989, bib:Zaritsky1994, bib:KobulnickyZaritsky1999, bib:MelbourneSalzer2002, bib:Lamareille2004, bib:Salzer2005a, bib:Lee2006, bib:Nagao2006, bib:Ellison2008, bib:KewleyEllison2008, bib:Guseva2009, bib:Berg2012, bib:Shi2014, bib:Zahid2014, bib:GonzalesDelgado2014, bib:Maier2015}).
In addition, much recent work has been completed examining the apparent evolution of such relationships with increasing redshift (e.g., \citealp{bib:Lara-Lopez2009, bib:Lara-Lopez2010, bib:Zahid2011, bib:Moustakas2011, bib:Zahid2013, bib:Lara-Lopez2013b, bib:Yabe2014, bib:Wuyts2014, bib:Izotov2015, bib:Wuyts2016}).
The failure thus far to arrive at universal consensus regarding the shape and structure of galaxian metallicity relations both locally and at high redshift speaks to such studies' inherent difficulties, necessitating continued examination in order to more fully understand the astrophysics associated with chemical evolution in star-forming systems.
\\
\indent The luminosity-metallicity (\LZ) relation compares gas-phase abundance with the constituent stellar population, traced by its measured absolute magnitude.
More luminous galaxies are generally found to be more metal-rich.
The mass-metallicity (\MZ) relation similarly compares gas-phase abundance with the constituent stellar population, traced instead by estimations of the stellar mass $M_{*}$.
Stellar mass is a more direct representation of a given system's stellar content than luminosity, but the determination of this parameter is not a direct observable and is therefore more difficult to ascertain reliably.
Calculations of $M_{*}$ are achieved through spectral energy distribution (SED) fits to photometric or spectroscopic data or through simpler mass-to-light conversions.
Recent achievements in SED fitting techniques provide results that are quite robust \citep{bib:Walcher2011}.
More massive galaxies are typically found to be more chemically enriched than less massive systems.
\\
\indent Two primary explanations exist to explain the observed correlation between stellar content and metallicity.
The first is that, due to the presence of a deeper gravitational potential well, massive galaxies are more capable of retaining their gas, and therefore their measured metallicities increase as compared to low-mass galaxies, which remain comparatively more metal-poor.
The second explanation is that there is a dependence of the star-formation efficiency upon the mass of the system such that massive galaxies are more effective at converting their gas into stars (and therefore producing new metals by stellar nucleosynthesis) than low-mass systems which remain unenriched over longer time scales.
Dwarf galaxies are thus predicted to have lower measured metallicities than higher-mass systems.
Simulations modeling only this mass-dependent star-formation efficiency, with no invocation of gas infall or outflow, have successfully reproduced the \MZ\ relation, implying that this latter process is the dominant physical motivation \citep{bib:Calura2009}.
The existence of actively star-forming dwarf galaxies (e.g., blue compact dwarfs) that are very efficiently forming stars yet possess low metal abundances, however, are a counter-example for this scenario.
\\
\indent Recently, studies have begun exploring the possibility of including a third parameter to these metallicity relations in an effort to reduce the intrinsic scatter (e.g., \citealp{bib:Lara-Lopez2010, bib:Mannucci2010, bib:Mannucci2011, bib:Yates2012, bib:Lara-Lopez2013b, bib:AndrewsMartini2013, bib:Lara-Lopez2013a, bib:Bothwell2013, bib:Stott2013, bib:NakajimaOuchi2014, bib:Maier2014, bib:Salim2014, bib:Gronnow2015, bib:DeRossi2015, bib:Jimmy2015, bib:Wu2016, bib:Brown2016}).
\citet{bib:Ellison2008} were the first to note that galaxies of a given mass with a higher star-formation rate (SFR) are systematically more metal-poor than those with a lower SFR.
Development of a three-dimensional fundamental plane (FP) relating stellar mass, gas-phase metallicity, and star-formation rate has been thus far successful in reducing scatter in the relationships between the individual parameters.
Furthermore, the $M_{*}$--$Z$--SFR relation appears to be redshift-independent \citep{bib:Mannucci2010}.
\\
\indent The functional forms of the \LZ, \MZ, and $M_{*}$--$Z$--SFR relations as determined by the various studies available in the literature demonstrate variations, attributable to such factors as differences in sample selection and the method used for deriving the metallicity.
A fit made to, for example, low-metallicity, high-excitation objects is restricted in its applicability to any but similar targets.
Galaxies of a more moderate degree of enrichment or star-formation activity, then, would potentially be poorly represented by such a fit.
Making use of the statistical power of large datasets such as SDSS helps to mitigate such limitations, but can still be biased by the large-scale effects influencing the overall sample.
Universality is thus a difficult characteristic for any such metallicity relation to claim.
\\
\indent A statistically complete sample of SFGs coupled with self-consistent oxygen abundance and stellar mass determination methods should provide a superior galaxy catalog for the derivation of representative metallicity relations of galaxies in the local universe.
The KPNO International Spectroscopic Survey (KISS; \citealp{bib:Salzer2000, bib:Salzer2001, bib:Salzer2002, bib:Gronwall2004b, bib:Jangren2005}) has identified over 2500 emission-line galaxies (ELGs) with absolute magnitudes spanning from $M_{B}$ = -22 to $M_{B}$ = -12.
It includes a range of object types including massive starburst galaxies, intermediate-mass irregular galaxies, low-mass dwarf irregulars, and blue compact dwarfs (BCDs) amongst its constituents.
The development of metallicity relations utilizing KISS star-forming systems provides an important comparison to the myriad studies available in the literature.
\\
\indent The current paper is part of a series of studies based on the KISS ELG sample that explore the metallicities of the survey constituents, either in small subsets (e.g., \citealp{bib:Melbourne2004, bib:Lee2004, bib:Hirschauer2015}) or as an ensemble (e.g., \citealp{bib:MelbourneSalzer2002, bib:Salzer2005b}).
The primary goal of this work is to expand on the latter two papers to create updated metallicity relations for the KISS sample.
In particular, we utilize new spectral data to greatly expand the size of our sample of KISS SFGs that possess metallicity estimates.
Furthermore, we are able to apply newly derived stellar masses for the first time to create \MZ\ and \MZSFR\ relations for the KISS galaxies.
The comparison of our updated metallicity relations with those from the literature results in the identification of possible biases present in previous metallicity relation studies.
\\
\indent In \S2 we discuss the KISS sample, our work to expand the dataset, and the properties of the galaxies selected from it.
Section 3 details the methods used to discern metallicities for the galaxies of this sample, and includes an exploration of the re-calibration of the strong emission line (SEL) empirical method used to create a self-consistent scale for oxygen abundances.
In \S4 we present the metallicity relations developed using our sample and abundance determination methods.
In \S5 we compare these relations with other studies available in the literature and discuss the implications our findings have on aspects of chemical evolution and enrichment histories.
Section 6 summarizes out results, and a description of our stellar mass determination method is presented in the appendices.
Throughout this work, we assume a standard cosmology of $\Omega_{\Lambda}$ = 0.73, $\Omega_{M}$ = 0.27, and H$_{0}$ = 70 km s$^{-1}$ Mpc$^{-1}$.

\section{The Data} 

\subsection{The KPNO International Spectroscopic Survey} 

\indent The KISS project was a wide-field objective-prism survey undertaken using the Burrell Schmidt 0.61-m telescope at Kitt Peak National Observatory (KPNO).
It was the first purely digital objective-prism survey for emission-line galaxies, identifying objects primarily through detection of \HA\ emission.
In addition, one survey strip cataloged galaxies detected via \OIII$\lambda$5007 emission.
The goal of the KISS project was to observe a large area of the sky for extragalactic emission-line sources, reaching a minimum of 2 magnitudes deeper than any previous photographic line-selected Schmidt survey.
KISS attempts to address scientific questions which require large samples with well-defined selection criteria and completeness limits.
In total, KISS catalogued over 2500 emission-line sources in four survey lists \citep{bib:Salzer2000, bib:Salzer2001, bib:Salzer2002, bib:Gronwall2004b, bib:Jangren2005}.
\\
\indent KISS constitutes a statistically complete, emission-line flux-limited sample of star-forming galaxies and active galactic nuclei (AGN).
The completeness characteristics of the survey are described in the individual survey papers.
The primary ELG sample is limited in redshift to $z$~=~0--0.095, and has a limiting line flux of $\sim$1.0 $\times$ 10$^{-15}$ erg s$^{-1}$ cm$^{-2}$.
This redshift range was achieved using a filter restricting detection of \HA\ emission to between 6400 and 7200 \AA\ for the \HA-selected portion of the survey.
The single \OIII-selected list \citep{bib:Salzer2002} similarly utilized a filter restricting the wavelength coverage of emission to between 4800 and 5500 \AA.
There is significant overlap of the \OIII-selected KISS catalog in terms of sky coverage and object detection with the first \HA\ survey list \citep{bib:Salzer2001}.
Only 91 objects were detected using the \OIII\ line alone, and most of these are found in regions where the two surveys did not overlap.
As a line flux-limited survey, KISS is able to sample a representative population of star-forming galaxies and AGN in the local universe.
Dependance upon detection of \HA\ emission for cataloging objects rather than, for example, solely upon \OIII$\lambda$5007 emission (e.g., for studies isolating objects with high-excitation spectra at low redshifts), allows KISS to remain largely unbiased to metallicity effects inherent to the reliance upon emission from metal lines.
KISS therefore remains sensitive to galaxies of all metallicities, even those of the highest and lowest abundances, when \OIII\ becomes weaker.
\\
\indent While the original objective-prism spectra of the galaxies detected by KISS were adequate to determine the presence of emission-lines and an estimate of the redshift, followup observations were necessary to confirm the specific source of spectral activity (i.e., star-forming galaxies versus AGN).
A campaign of ``quick-look" spectroscopy has been undertaken by members of the KISS group on a variety of observational facilities, including the 9.2-m Hobby-Eberly Telescope (HET; \citealp{bib:Gronwall2004a}), the Lick 3-m telescope \citep{bib:Melbourne2004}, the MDM 2.4-m telescope \citep{bib:Wegner2003, bib:Jangren2005}, the WIYN 3.5-m telescope \citep{bib:Salzer2005b}, the ARC 3.5-m telescope, the KPNO 2.1-m telescope, the KPNO 4-m telescope, and the Keck I 10-m telescope \citep{bib:Hirschauer2015}.
The multitude of spectra span a wide range in signal-to-noise (S/N) ratios and wavelength coverage regimes.
The full optical spectrum, ranging from blueward of \OII$\lambda\lambda$3726,3729 and redward of \SII$\lambda\lambda$6716,6731, is not observed for all KISS objects due to the limited wavelength coverage of some of the spectrographs used.
All available KISS spectra include at a minimum the emission lines of \HA, \HB, \OIII$\lambda\lambda$4959,5007, and \NII$\lambda\lambda$6548,6583, which are generally required for the derivation of SEL empirical abundance estimations.

\subsection{Updating the KISS Database} 

\begin{figure} 
\includegraphics[height=8.6cm,width=8.6cm]{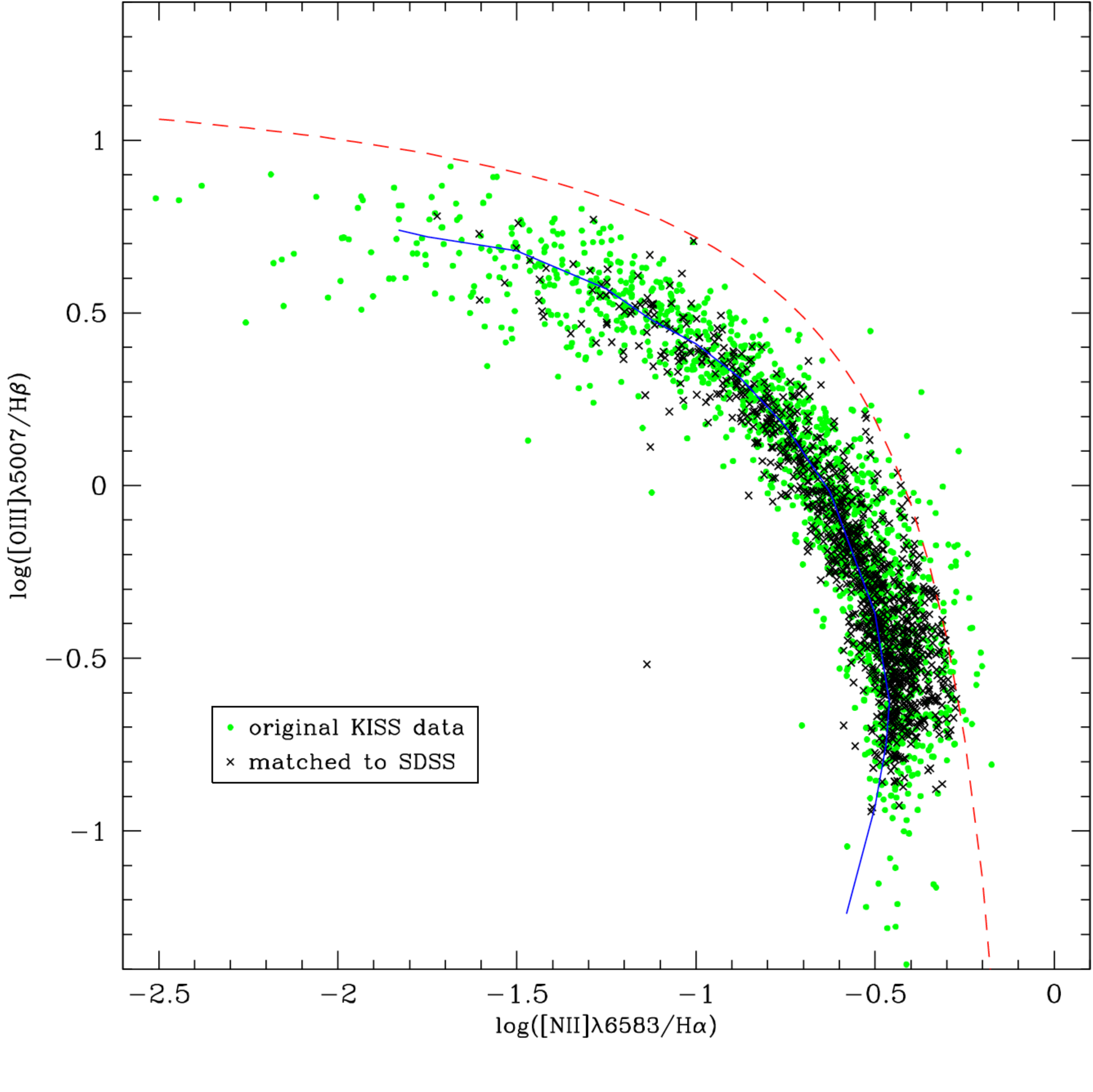}
\caption{\footnotesize Spectral activity diagnostic diagram of star-forming galaxies from KISS (green circles) overlaid with targets matched to SDSS (black crosses).
Matched objects' spectral information was compared to existing KISS values and used to supplement the database in the case of missing or spurious data.
A large number of matched galaxies reside in the lower-right side of this diagram, consistent with relatively high-abundance sources.
In many cases, emission line data for \OII$\lambda$3727 was added, critical for deriving accurate oxygen abundances.
The solid blue line represents a star-formation sequence from theoretical models \citep{bib:DopitaEvans1986}, while the dashed red line is an empirically-determined demarcation between SFGs and AGN \citep{bib:Kauffmann2003}.
}
\label{fig:KISS_SDSS_DD}
\end{figure}

\indent At the time of the most recent presentation of the KISS sample for calibration of an SEL abundance method and formulation of an \LZ\ relation \citep{bib:Salzer2005a}, 1,351 KISS galaxies had follow-up spectral data.
Subsequent updates by the KISS group have increased that number to 2,464.
These new spectra have been obtained from observations using the 9.2-m HET, 2.4-m MDM, 2.1-m KPNO, and 3-m Lick telescopes.
As of this writing, 100\% of the KISS galaxies in the \HA-selected survey lists 1 and 2 \citep{bib:Salzer2001, bib:Gronwall2004b}, the \OIII-selected list \citep{bib:Salzer2002}, and the Spring region of the third \HA-selected list \citep{bib:Jangren2005} have follow-up spectra.
A subsequent data paper will present these more recent observations (Salzer et al., {\it in preparation}).
%
%
\\
\indent The majority of the new observations were undertaken using HET.
The spectrograph setup utilized covered the wavelength range $\sim$4350--7250 \AA.
This precluded the observation of the \OII$\lambda\lambda$3726,3729 doublet and, in most cases, the auroral \OIII$\lambda$4363 line.
Due to this, the HET spectra could not be used to derive ``direct-method" metallicities or to utilize any of the \R-based abundance methods (e.g., \citealp{bib:McGaugh1991}).
Observations using the MDM and Lick telescopes, however, utilized spectrographs with sufficient short-wavelength sensitivity to include measurement of the strong \OII\ emission line doublet.
In addition, some followup observations recovered \OIII$\lambda$4363, allowing for computation of the electron gas temperature (\Te) and thus determination of direct-method abundances (e.g., \citealp{bib:Hirschauer2015}).
\\
\indent To complement the observations undertaken by members of the KISS group described above, we have additionally matched KISS galaxies to SDSS targets in an effort to further supplement the database (see \citealp{bib:Strauss2002} for a description of SDSS data products).
This is illustrated using a standard spectral activity diagnostic diagram (e.g., \citealp{bib:Baldwin1981}), showing the location of KISS galaxies (green circles) overlaid with targets matched to SDSS (black crosses) in Figure \ref{fig:KISS_SDSS_DD}.
In total, 1,121 KISS galaxies were observed spectroscopically by SDSS, constituting $\sim$44\%\ of the total sample.
Because the median $m_{B}$ of 18.1 (e.g., \citealp{bib:Salzer2001, bib:Gronwall2004b}) for the KISS sample is comparable to the spectroscopic limit for SDSS, the overlap is far less than 100\%.
For any KISS galaxy lacking followup spectroscopy, we have adopted SDSS spectral data when available.
A comparison of line flux ratios for galaxies observed by both surveys indicates that the two datasets generally exhibit good agreement, allowing for the adoption of SDSS data when necessary.
The addition of the SDSS spectroscopy has yielded usable emission-line data for an additional 319 galaxies that had otherwise not yet been observed by members of the KISS group.
We note that as the SDSS spectral coverage range is blue-sensitive enough to include \OII$\lambda$3727 for galaxies with redshifts above $z$ $\sim$ 0.03, these supplemental data have been used to compute \R-based metallicities for applicable KISS objects.
\\
\indent For cases in which a matched KISS galaxy possessed existing spectral data but lacked a measurement of \OII, we adopted the SDSS emission line flux information for this doublet.
This procedure required confirmation that these two distinct datasets were measuring similar emission regions.
While the KISS observing strategy attempts to place the spectrograph slit over the region of the galaxy with strongest emission as detected in the objective-prism spectrum, the automated SDSS system generally selects the brightest location in the broad-band image (see \citealp{bib:Strauss2002} for details).
Observations of discrepant physical regions may become manifest as inconsistencies in the measured emission line ratios of the nebular gas.
Furthermore, the slit-widths of the KISS long-slit spectroscopy are generally 1.0-1.5 arcseconds, while SDSS fibers used on KISS galaxies are 3.0 arcseconds in diameter.
%
%
These differences in aperture necessitate that, even when the target object is the same for both, the observed regions of the KISS and SDSS spectra are not identical.
In order to identify mismatches in the two datasets' spectra, we have compared the galaxies' \HB\ equivalent widths (EW) and \OIII/\HB\ flux ratios, rejecting the SDSS \OII\ value for deviations of either by more than 40\% from equality.
This 40\% departure limit corresponds roughly to a 2$\sigma$ rejection limit.
In total, 239 KISS objects lacking previous \OII\ observations had their spectral data augmented with the addition of \OII\ from SDSS, representing $\sim$10\% of the total number of galaxies within the KISS sample.
\\
\indent In total, an additional 1,113 new KISS galaxy spectra were added to the database since the \citet{bib:Salzer2005a} analysis, thanks to the additional observations undertaken subsequent to the previous KISS metallicity relations study and cross-matching objects with SDSS.
Many of these new spectra include \OII$\lambda$3727 emission that is crucial for SEL estimations of oxygen abundance, and some include the temperature-sensitive \OIII$\lambda$4363 line necessary for robust direct-method metallicity determinations.
This allows for a substantial increase in the number of calibration points used for the development of an SEL empirical abundance method applicable to all of the galaxies in the sample.
While the previous study of \citet{bib:Salzer2005a} included 185 galaxy spectra with both the strong \OIII\ and \OII\ lines necessary for the development of \R-method metallicities used for SEL method calibration, this study includes 739 galaxies' spectra used for such a calibration, representing a nearly four-fold increase.
In particular, we possess a substantial increase in \R-based abundances for higher-metallicity systems.

\subsection{KISS Stellar Mass Determination} 

Determination of the total stellar mass for galaxies in the KISS database was accomplished using fits of spectral energy distributions (SEDs) to photometric data obtained from a variety of sources.
These masses are utilized for the derivation of \MZ\ and \MZSFR\ FP relationships.
The use of multi-wavelength SED-fitting techniques was deemed essential for the KISS galaxies, since simpler methods involving mass-to-light ratio conversions (e.g., \citealp{bib:BelldeJong2001}) are subject to large uncertainties when applied to galaxies dominated by starbursts.

\subsubsection{Photometric Data} 

The input photometric data to construct SEDs for the KISS objects come from a variety of archival data sources, including:
Optical fluxes from the KISS database \citep{bib:Salzer2000, bib:Salzer2001}, SDSS Data Release 12 (DR12; \citealp{bib:Alam2015}), and the NASA Sloan Atlas (e.g., \citealp{bib:Blanton2011}); ultraviolet (UV) fluxes from the \emph{Galaxy Evolution Explorer} (\emph{GALEX}; \citealp{bib:Martin2005, bib:Morrissey2007}); near-infrared (NIR) fluxes from the Two Micron All Sky Survey (2MASS; \citealp{bib:Skrutskie2006}); and mid-infrared (MIR) fluxes from the \emph{Widefield Infrared Survey Explorer} (\emph{WISE}; \citealp{bib:Wright2010}).
\\
\indent Appendix~A includes a detailed description of the photometric data, including the verifications and quality tests we used to determine which photometric measurements to include in our SED fits.

\subsubsection{SED-fitting Grids} 

SEDs for each KISS galaxy were fit using the Code Investigating Galaxy Emission (CIGALE) software \citep{bib:Noll2009}.
CIGALE creates a grid of synthetic SEDs based on theoretical models and determines the best parameter values by comparing with our observed SEDs.
Our multi-wavelength data (from UV to IR) allows CIGALE to account for dust absorption and re-emission in a self-consistent manner.
Nebular emission and absorption is included from the templates of \citet{bib:Inoue2011} which are based on the amount of ionizing UV flux (see also Section 5.5 of \citealp{bib:Salim2016}).
Dust attenuation is based on the method of \citet{bib:Cardelli1989}, with the formulas from \citet{bib:Calzetti2000} and \citet{bib:Leitherer2002}.
Our fitting process follows the methodology developed in \citet{bib:Janowiecki2017a}, and is described in more detail in Appendix~B.
\\
\indent We have derived reliable stellar masses using this SED fitting process for 2,207 KISS galaxies.
For most objects ($N$ = 1,946), SED fits were computed utilizing some combination of UV, optical and IR data, and for all but 67 systems we have at least optical and infrared data.
The near-infrared flux is particularly important for determining the stellar mass as it acts as a tracer for the older stellar population of a given system.
Ideally, an SED fit is populated with photometric data from all wavebands.
In some circumstances, however, these data are not available.
Disparities in coverage area and/or depth, for example, may preclude the inclusion of a given KISS galaxy into one or more photometric surveys.
We fully describe these inhomogeneities in Appendix~B.

\section{KISS Galaxy Metallicities} 

\indent Global metallicities are ascertained through a variety of means.
As described in the previous section, the optical spectra for KISS galaxies have been obtained using a diverse assortment of telescopes and instruments with a broad variety of wavelength ranges, resolutions, and dispersions.
We describe here how this diverse spectral dataset is used to derive self-consistent oxygen abundances for all star-forming galaxies in the survey with spectra of sufficient quality.
Direct-method abundances were calculated whenever possible, but only a small fraction of KISS galaxies possess suitable data, so strong-line empirical relations must be employed for the remaining sources.
Well-defined SEL techniques such as the \citet{bib:McGaugh1991} abundance grid method provide a robust metallicity estimate for many KISS galaxies that lack \Te-method abundances.
The combination of direct- and McGaugh-method abundances provide the calibration points required for development of a single empirical relation that is applicable to all objects in the KISS database, many of which lack metallicities from other methods.
These abundance determination techniques are described in greater detail in the following subsections.
\\
\indent For the analysis carried out in the remainder of this paper, we utilize only those KISS ELGs whose spectral data are classified as having quality code 1 or 2 (see the various spectral data papers listed above) and are classified as SFGs.
This ensures that the emission-line ratios used for this analysis are reliable (i.e., have uncertainties of $\sim$0.1 dex or better).
Overall, this results in a total of 1,468 star-forming KISS ELGs.

\subsection{Direct-Method Metallicities} 

\indent Direct-method abundances are the preferred metallicity determination method available for star-forming systems, requiring measurement of the electron density \Ne\ and the electron temperature \Te.
For our KISS galaxy spectra, \Ne\ is calculated from the \SII$\lambda$6716/$\lambda$6731 ratio, and is typically found to be $\sim$100 e$^{-}$ cm$^{-3}$.
\Te\ is determined by the line ratio of doubly-ionized oxygen \OIII$\lambda$4363/\OIII$\lambda\lambda$4959,5007 \citep{bib:OsterbrockFerland2006}.
The temperature of singly-ionized oxygen is estimated by using the algorithm presented in \citet{bib:Skillman1994} based on the nebular models of \citet{bib:Stasinska1990},
\[
t_{e}(\text{O}^{+}) = 2[t_{e}(\text{O}^{++})^{-1} + 0.8]^{-1},
\]
\noindent where $t_{e}$ are temperatures measured in units of 10$^{4}$ K.
The total oxygen abundance is assumed to be given by
\[
\frac{\text{O}}{\text{H}} = \frac{\text{O}^{+}}{\text{H}^{+}} + \frac{\text{O}^{++}}{\text{H}^{+}},
\]
following the standard practice.
For a more thorough explanation of our treatment of \HII\ region diagnostics, see \S2.5 of \citet{bib:Hirschauer2015}.
\\
\indent For star-forming galaxy spectra with sufficient sensitivity and resolution, the detection and measurement of all aforementioned emission lines required for direct-method metallicity analysis is possible.
Systems with strong emission lines are generally those possessing higher \Te, and are therefore often those of lower oxygen abundance.
In total, 71 KISS ELGs possess \OIII$\lambda$4363 detection.
The \citet{bib:Salzer2005a} study included only 22 objects with direct-method abundances, adopted from the previous KISS metallicity papers of \citet{bib:Melbourne2004} and \citet{bib:Lee2004}.
An additional 49 KISS galaxies with \Te-method abundances were computed using spectra taken primarily at the Lick and MDM Observatories by members of the KISS team.
This represents a roughly three-fold increase in the total number of robust oxygen abundances included within the KISS database.
Calculations of physical parameters (e.g., \Te) of the ionized gas as well as the oxygen abundance were performed using the Emission Line Spectrum Analyzer (\texttt{ELSA}) program \citep{bib:Johnson2006} as described in detail in \citet{bib:Hirschauer2015}.

\subsection{McGaugh Grid Metallicities} 

\begin{figure} 
\includegraphics[height=8.6cm,width=8.6cm]{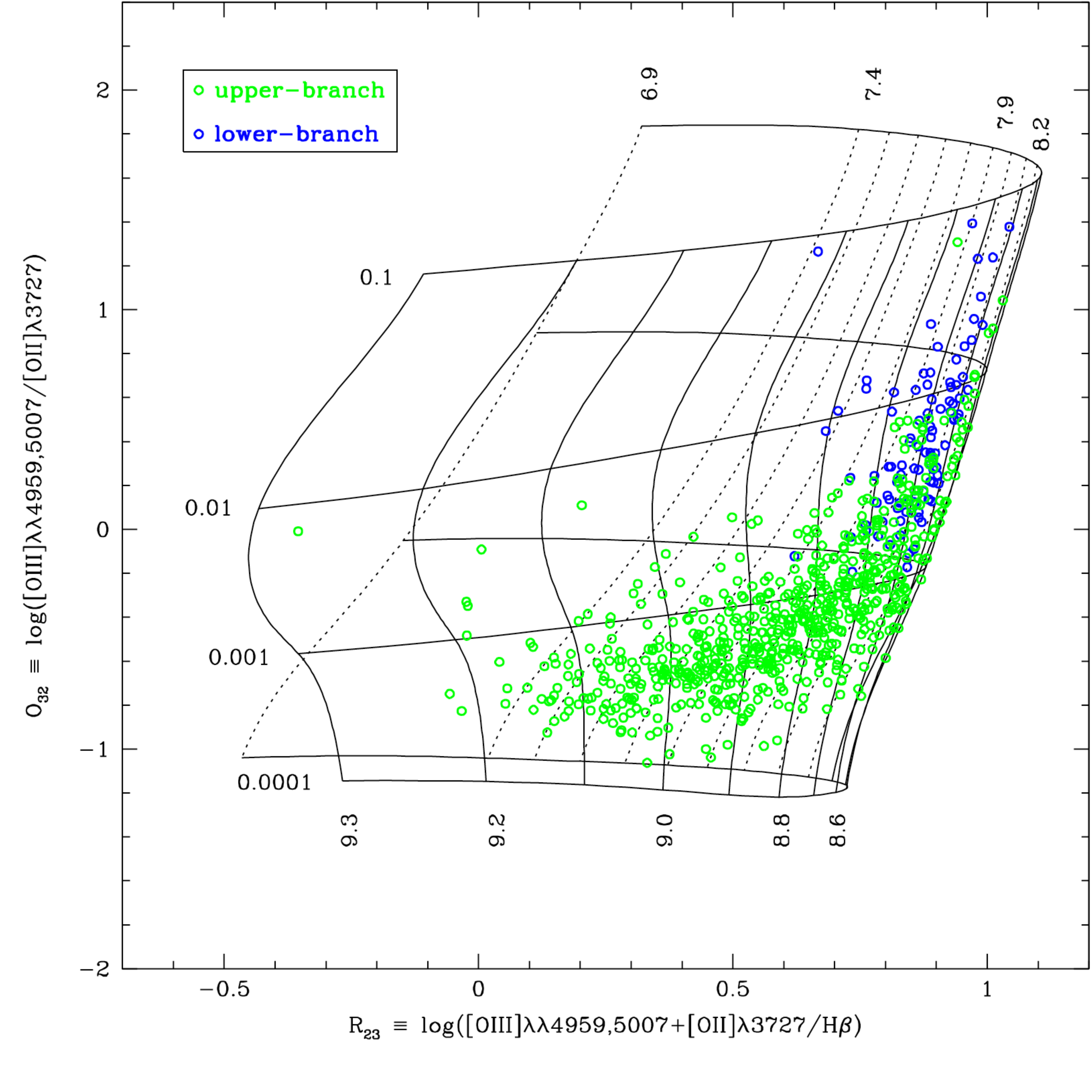}
\caption{McGaugh abundance grid \citep{bib:McGaugh1991} for galaxies in the KISS sample.
In total, metallicities were computed using the McGaugh-method for 756 objects.
Galaxies on the upper-metallicity branch (foreground) are indicated with open green circles, while galaxies on the lower-metallicity branch (background) are indicated with open blue circles.
Demarcation between branches is determined using the abundance-sensitive \NII/\OII\ line ratio \citep{bib:vanZee1998}.
}
\label{fig:KISS_McGaugh}
\end{figure}

\indent For cases where \Te-method abundances are not obtainable, metallicities are estimated through techniques employing bright lines.
These are either empirical relations that are calibrated to direct-method abundances, theoretical relations that are based on photoionization models, or semi-empirical relations that are a mixture of the two (e.g., \citealp{bib:Moustakas2010}).
Bright lines used for such SEL methods are consistently measurable in most spectra, making these techniques useful for a much larger fraction of objects than the direct method, however they necessarily possess greater intrinsic scatter.
The most common SEL method is \R\ $\equiv$ log((\OII$\lambda\lambda$3726,3729+\OIII$\lambda\lambda$4959,5007)/\HB), first introduced by \citet{bib:Pagel1979}.
\R\ provides an estimate of total cooling and thus represents a proxy for global oxygen abundance \citep{bib:KewleyDopita2002}, but is famously double-valued at high- and low-abundance and includes an ambiguous turnaround region.
Many \R\ calibrations are available in the literature \citep{bib:EdmundsPagel1984, bib:McCall1985, bib:Skillman1989, bib:McGaugh1991, bib:Zaritsky1994, bib:Pilyugin2000, bib:Pilyugin2001, bib:KewleyDopita2002, bib:Denicolo2002, bib:PettiniPagel2004, bib:KobulnickyKewley2004, bib:PilyuginThuan2005}.
\\
\indent The \citet{bib:McGaugh1991} model grid is an improvement on \R, derived using photoionization model calculations from \texttt{CLOUDY} \citep{bib:FerlandTruran1981} and employing the abundance-sensitive \R\ line ratio alongside the excitation-sensitive $O_{32}$ line ratio, defined as log(\OIII$\lambda\lambda$4959,5007/\OII$\lambda$3727).
It incorporates an upper- and lower-metallicity branch along the axis defined by the \R\ line ratio.
Typical abundance uncertainties of the McGaugh model grid method are $\sigma$ $\approx$ 0.05 for the low-metallicity branch and $\sigma$ $\approx$ 0.10 for the high-metallicity branch, dominated by the uncertainties in the model stellar radiation fields \citep{bib:McGaugh1991}.
Abundances were computed by means of the McGaugh model grid for every KISS galaxy possessing the requisite emission line data in order to provide reliable metallicity information when \Te-method calculations were not possible.
Following \citet{bib:vanZee1998}, we employ the \NII/\OII\ line ratio to differentiate between objects on the upper and lower \R\ branch.
In total, McGaugh-method metallicities were computed for 756 KISS galaxies, as illustrated in Figure \ref{fig:KISS_McGaugh}, comprising 38\% of the total number of SFGs in the database.
The majority of these abundances were calculated for galaxies on the upper-metallicity branch (foreground; open green circles).
Lower-metallicity branch galaxies (background; open blue circles) exhibit noticeably higher excitations.
\\
\indent We note that some recent studies have discovered a modest systematic offset between direct- and McGaugh-method metallicities, typically in the range of  0.10-0.25 dex (e.g., \citealp{bib:Hirschauer2015, bib:Bresolin2009}; see also discussion in \citealp{bib:KewleyEllison2008}).
Of the 71 KISS star-forming galaxies with calculated \Te-method abundances, 58 also have metallicities computed using the McGaugh grid (the remaining 13 objects have line-ratios which fell off of the McGaugh abundance grid, and so McGaugh-method metallicities were not computed; these objects are located in the so-called turnaround region).
For these systems, the mean offset in abundance is $\Delta$(log(O/H))~=~0.11 dex in the sense that the direct abundances are lower.
Many (23) of these galaxies reside in the ambiguous turnaround region of the McGaugh abundance grid, and so for the remaining 35 systems this mean abundance offset is reduced to $\Delta$(log(O/H))~=~0.09 dex.
Several possible explanations exist for this abundance offset, including, for example, small-scale electron temperature fluctuations (e.g., \citealp{bib:Peimbert1967, bib:PeimbertCostero1969, bib:Peimbert2003, bib:Peimbert2007}), thermal gradients \citep{bib:Garnett1992}, overestimation of O$^{+}$ zone temperatures compared to O$^{++}$ zone temperatures \citep{bib:AndrewsMartini2013}, and discrepancies in models of electron energy distributions (the ``$\kappa$-distribution"; see \citealp{bib:Nicholls2012, bib:Nicholls2013, bib:Dopita2013}).
For a more complete discussion, we refer to \S2.6.1 of \citet{bib:Hirschauer2015}.
The cause of the discrepancy between empirically derived direct abundances and abundances derived from photoionization models remains unclear.  
For objects in the current study the average metallicity difference is comparable to the formal uncertainties associated with the McGaugh method. 
We consider the impact of the modest offset between our direct and McGaugh method abundances on the derivation of our SEL metallicity relation in the next section.

\subsection{O3N2-Method Calibration} 

\indent Because the \OII$\lambda$3726,3729 doublet was not observed for many KISS galaxies due to limitations of the spectrographs used, a large number of KISS star-forming galaxies lacked the required information needed to compute metallicities via the McGaugh model grid described in \S3.2.
Some other method, then, was necessary to derive metallicities for {\it all} KISS objects with suitable-quality spectra.
\\
\indent The Coarse abundance method pioneered by \citet{bib:MelbourneSalzer2002} uses the line ratios \NII$\lambda$6583/\HA\ and \OIII$\lambda$5007/\HB\ since they are observed in essentially all KISS ELG spectra, typically have good S/N ratios, and are both fairly insensitive to reddening corrections.
A standard spectral activity diagnostic diagram (\citealp{bib:Baldwin1981}; see Figure \ref{fig:KISS_SDSS_DD}) utilizes the logarithms of these same line ratios as the horizontal and vertical axes, respectively.
In such a plot, metallicity varies continuously over the distribution of galaxies, with low-metallicity systems in the upper-left and high-metallicity galaxies in the lower-right.
At the low-metallicity end, the \OIII$\lambda$5007/\HB\ line ratio remains almost constant for a large range of $\log$(\NII$\lambda$6583/\HA) values.
Similarly, on the high-metallicity end, the \NII$\lambda$6583/\HA\ line ratio stays nearly constant for a large range of $\log$(\OIII$\lambda$5007/\HB) values.
The correlation between these two line ratios and oxygen abundance (or a combination of the the two in the intermediate regime) can therefore be used as an estimate for metallicity when calibrated to sources of known abundance.
\\
\indent Calibrations of the Coarse abundance method were previously developed by \citet{bib:MelbourneSalzer2002} and later refined by \citet{bib:Salzer2005a}, utilizing the \Te- and McGaugh-method metallicities that were available for a small number of KISS galaxies at that time.
These calibrations were used to estimate the metallicities for the rest of the objects in the KISS sample to an uncertainty of $\sigma$ $\approx$ 0.20 dex.
While not exceedingly accurate, particularly in comparison to direct-method abundances, the Coarse method provides a useful means to study the relative distribution of oxygen abundances for galaxies in a given sample from a statistical perspective.
\\
\indent One challenge that had presented itself with the previous Coarse abundance method, however, is the inevitable discontinuity that arises in the transition zone between the $\log$(\NII$\lambda$6583/\HA)- and $\log$(\OIII$\lambda$5007/\HB)-sensitive regimes on the standard spectral activity diagnostic diagram.
In this region an average of the metallicities computed by the two relations is assumed.
While consistent within the errors of the method, discontinuities in the abundances computed in such a manner leaves artifacts in metallicity relations.
For example, there could be regions of parameter space whereby no combinations of measured emission-line ratios could produce an oxygen abundance estimate of certain values.
Clearly, this problem is undesirable, and as such led us to seek an alternative SEL technique.
\\
\indent The O3N2 $\equiv$ log\{\OIII$\lambda$5007/\HB)/(\NII$\lambda$6583/\HA)\} parameter, first introduced by \citet{bib:Alloin1979}, uses the same emission lines as the Coarse method but has the additional advantage of being fit by a single continuous function.
The functional form of this abundance estimator avoids the discontinuity present with the Coarse method, but maintains its overall utility.
Previous calibrations of the O3N2-method include the studies of \citet{bib:PettiniPagel2004}, \citet{bib:Nagao2006}, \citet{bib:Perez-MonteroContini2009}, and \citet{bib:Marino2013}.
These versions of O3N2 are typically evaluated against a set of ELGs and extra-galactic \HII\ regions, and are generally weighted toward the lower metallicity direct-method abundance objects available in the literature (which are definitively biased toward bright, higher-temperature, lower-metallicity objects) from which calibrations may be made.
Higher-metallicity systems are comparatively less well-represented by these O3N2 fits and as such the use of the available calibrations is not applicable in the high-abundance regime.
\\
\begin{figure} 
\includegraphics[height=8.6cm,width=8.6cm]{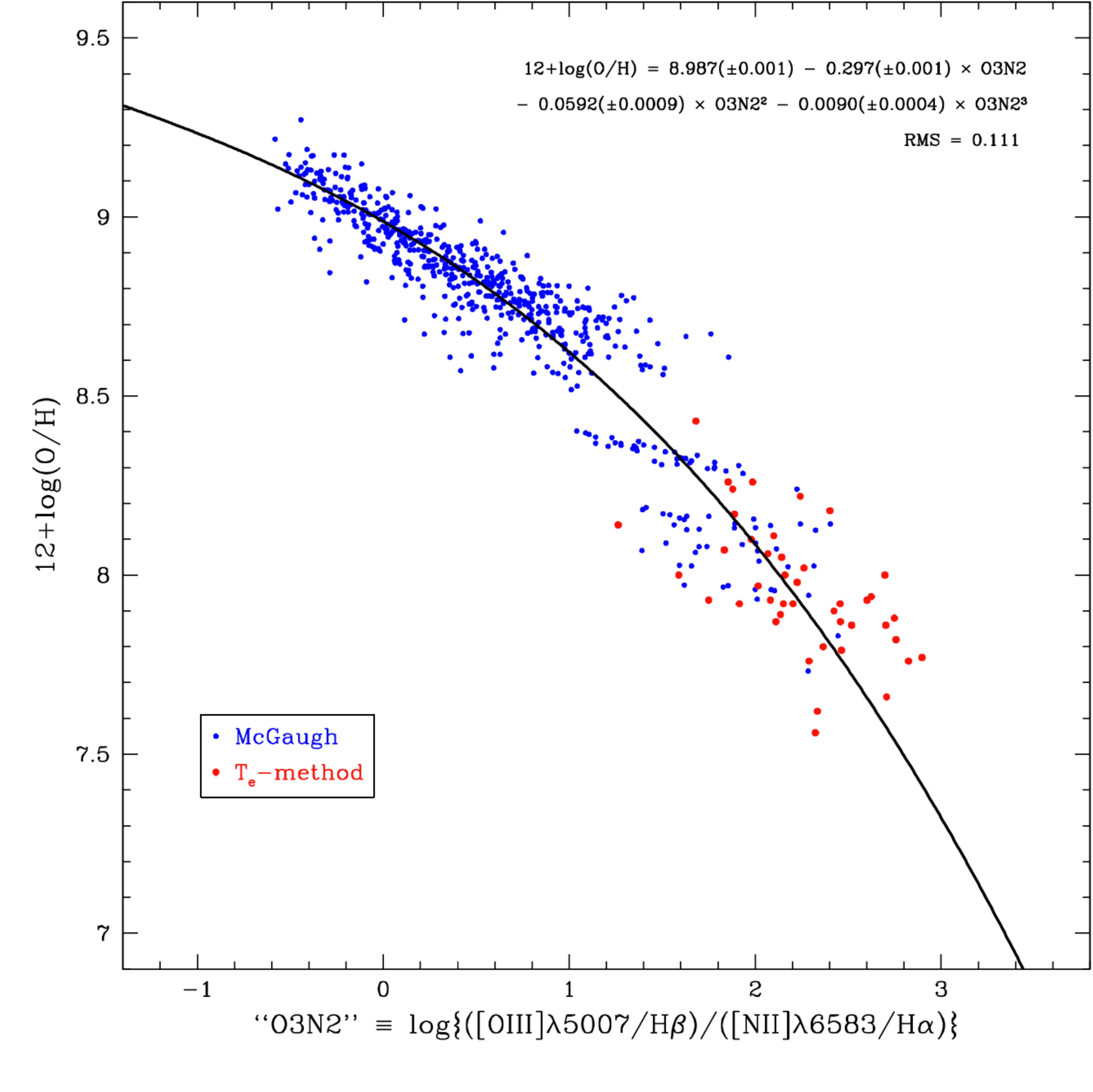}
\caption{Cubic bisector fit of the O3N2 metallicity index for KISS SFGs, defined as log\{\OIII$\lambda$5007/\HB)/(\NII$\lambda$6583/\HA)\}, versus oxygen abundance \abun.
Blue points are galaxies with metallicities derived by means of the McGaugh abundance grid, while red points are derived via the direct-method.
The O3N2 method has the advantage of utilizing the same set of readily-available strong emission line ratios as the coarse abundance method, but without the inconsistencies that arise as a consequence of the differing regions of applicability in the BPT diagram.
}
\label{fig:O3N2_calibration}
\end{figure}
\indent For this study, we have utilized our large sample of KISS galaxies that possess \Te- or McGaugh-method abundances to calibrate the O3N2 relation using data that spans both high and low abundances.
The O3N2 index was computed for our spectral data and compared to these metallicities as demonstrated in Figure \ref{fig:O3N2_calibration}.
McGaugh-grid metallicities are represented as blue dots, while \Te-method abundances are shown as red dots.
To find the functional form of this relation, we computed polynomial bisector functions to fit the data by varying the dependent and independent variables, then fitting a function to the average points.
We have found that a cubic polynomial minimizes the RMS scatter in the vertical (abundance) axis, resulting in a fit of the form:
\[
12+\log(\mathrm{O/H}) = 8.987(\pm0.001) - 0.297(\pm0.001) \times \mathrm{O3N2}
\]
\[
-\ 0.0592(\pm0.0009) \times \mathrm{O3N2}{^2} - 0.0090(\pm0.0004) \times \mathrm{O3N2}{^3},
\]
with RMS uncertainty of $\sigma$ = 0.111.
A comparison of Coarse- and O3N2-method metallicities for KISS SFGs demonstrates very little scatter.
We interpret this result as an indication of the interchangeability of the respective methods for this data set.
By being fit as a single continuous function, however, the O3N2-method has the advantage over the piecewise fit (including an intermediate zone utilizing an average value) of the Coarse method.
We therefore adopt the O3N2-method for the remainder of this paper for computing Coarse abundances for the KISS sample.
\\
\indent We note that there exists a general tendency toward lower oxygen abundances computed using direct-methods in comparison to SEL-methods (\S3.2; see also \citealp{bib:Hirschauer2015}).
The utility in adopting metallicities from both techniques in the calibration of a new empirical relation despite this offset may therefore seem questionable.
The robustness of \Te-method abundances makes them difficult to reject from our calibrations.
Alternatively, the limitations of deriving direct-method abundances for high-metallicity objects necessitates the use of alternative techniques to estimate their metallicities.
An application of both apparently discrepant methods is therefore borne out of necessity in order to account for galaxies of all metallicities.
With this necessity in mind and a conscious understanding of the fairly small offset between the \Te-method and the McGaugh abundance grid ($\sim$0.1 dex), we utilize both in our O3N2-method calibration.
In addition, because the cause of this abundance offset is not entirely clear, we refrain from applying a correcting offset (i.e., lowering SEL-derived metallicities to match direct-method metallicities or vice-versa) so as to not introduce systematic biases.
As a test, we developed a cubic polynomial fit replacing the \Te\ abundances with McGaugh-method abundances.
This produced a function consistent with that shown in Figure \ref{fig:O3N2_calibration}.
Differences in the two fits are minimal over the majority of the range covered by our galaxies, only approaching $\sim$0.1 dex at the extreme low-abundance end.

\section{Metallicity Relations} 

\indent The overall goal for this project was the development of a self-consistent abundance determination method that could be applied to the entirety of the KISS galaxy sample in order to produce new metallicity relations for local star-forming systems.
We believe that our calibration of the SEL O3N2-method metallicity indicator provides much improved determinations of the oxygen abundances for the KISS star-forming galaxies than what had been presented in the previous papers in the series.
%
%
The inclusion of significant numbers of McGaugh-method abundances for higher-metallicity targets has substantially increased the breadth of objects available for O3N2-method calibration compared to previous assessments.
In this section we utilize our O3N2-method calibration to derive updated metallicity relations for the KISS sample of ELGs.
Discussion of the resulting relations is deferred to \S5.

\subsection{Luminosity-Metallicity ($L$--$Z$)} 

\begin{figure} 
\includegraphics[height=8.6cm,width=8.6cm]{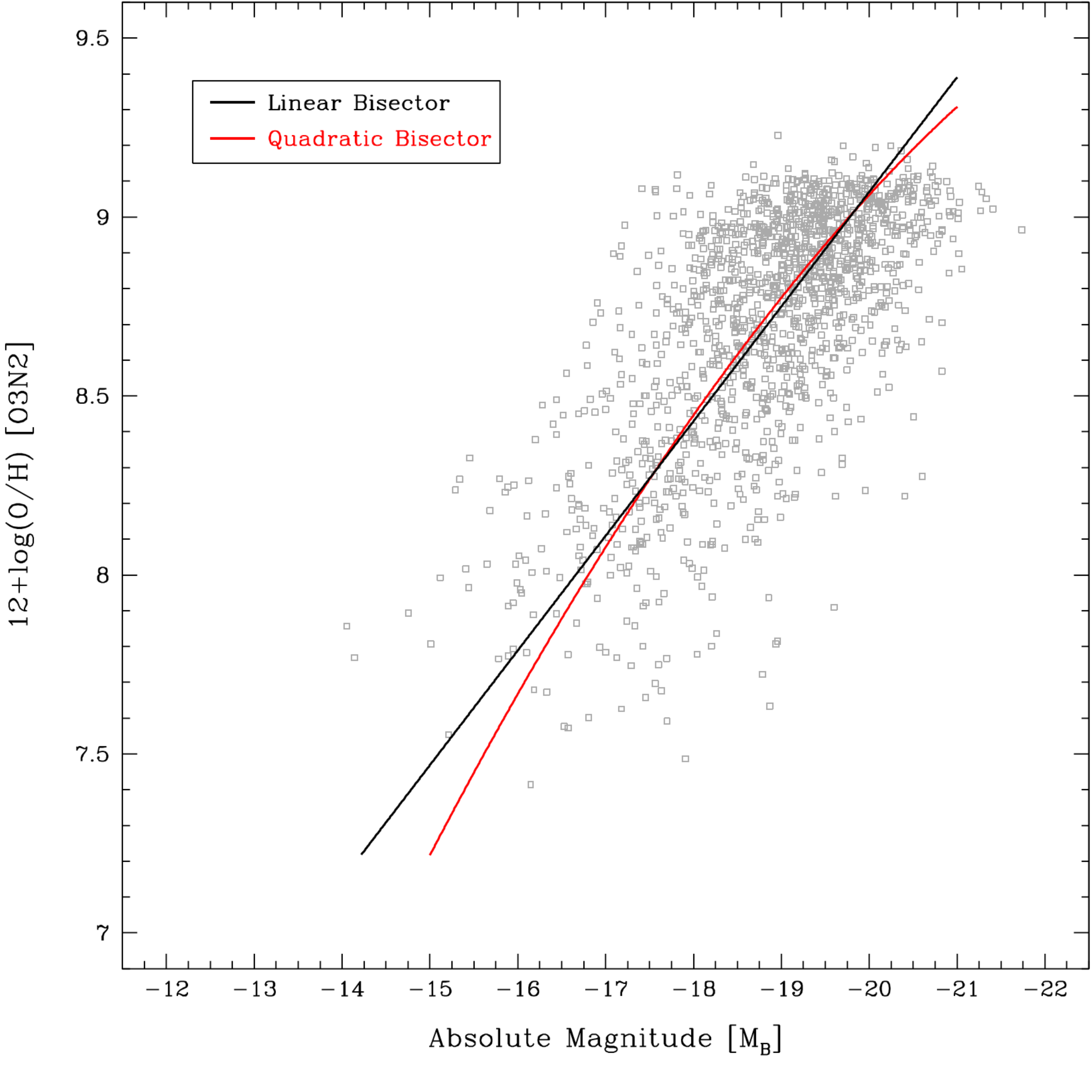}
\caption{Linear (black line) and quadratic (red line) bisector luminosity-metallicity (\LZ) relation fits to the KISS sample of star-forming galaxies.
Oxygen abundances for this sample are derived via the O3N2 abundance method, while luminosity is determined from KISS photometry.
The linear and quadratic fits are consistent with each other within the formal uncertainties over the range of $M_{B}$ where the majority of galaxy data lie.
}
\label{fig:LZ}
\end{figure}

\indent Luminous galaxies have consistently been found to be more metal-rich than low-luminosity galaxies.
The \LZ\ relationship has been shown to hold over $\sim$11 magnitudes in luminosity and $\sim$2 dex in metallicity, and does not appear to show dependence upon environment \citep{bib:Vilchez1995, bib:Wu2017} or morphology \citep{bib:Mateo1998}, an indication that the processes regulating \LZ\ are common to all galaxies.
Previous studies of \LZ\ have found that the slope of the relation is shallower for dwarf galaxies, however, implying that the linearity of the relationship is questionable over the full range of luminosity (e.g., \citealp{bib:MelbourneSalzer2002, bib:Tremonti2004, bib:Salzer2005a}).
Differences in the methods by which abundances are determined for high- and low-mass galaxies introduces difficulty in assessing the true shape of these relations.
\\
\indent Development of a consistent method for deriving metallicities for all galaxies was important for establishing metallicity relations that effectively span the full range of systems in the universe.
We have combined metallicity estimates using the O3N2 technique with absolute $B$-band magnitudes derived using KISS photometry (e.g., \citealp{bib:Salzer2000}) to investigate the form of the \LZ\ relation.
Apparent magnitudes collected from the imaging portion of the survey data have been corrected for Galactic reddening, and distances used to compute absolute magnitudes were derived from redshifts determined from the follow-up spectroscopy of each galaxy.
We note that 19 KISS galaxies with smaller redshift distances ($<$ 36.0 Mpc) have been excluded from our \LZ\ sample because of the large uncertainties implicit in using redshifts to determine distances for nearby objects where the peculiar velocities (i.e., non-Hubble flow velocity component) lead to large uncertainties in their luminosities and masses.
\\
\indent Figure \ref{fig:LZ} is an \LZ\ relation plot of KISS SFGs, with metallicities computed using the O3N2 method.
Derivation of the functional fit to the data was accomplished by finding the bivariate linear least-squares bisector of two separate linear fits.
Our fit of the KISS \LZ\ relation takes the following form:
\[
12+\log(\mathrm{O/H}) = 2.664(\pm0.170) - 0.320(\pm0.009) \times M_{B},
\]
where $M_{B}$ is the $B$-band absolute magnitude.
In total, 1,468 KISS SFGs are included in this plot.
The RMS scatter of the data about this fit is derived in the vertical (abundance) axis.
We find a scatter of $\sigma$ = 0.280.
\\
\indent The data were additionally fit with a quadratic bisector in effort to investigate the possible flattening of the \LZ\ distribution at high abundances as has been seen in some previous studies (e.g., \citealp{bib:Tremonti2004}).
Our \LZ\ relation does not show any strong flattening at high luminosity.
For completeness, we illustrate our quadratic fit to the data alongside our linear fit in Figure \ref{fig:LZ}.\footnote{\footnotesize Quadratic bisector fit polynomial coefficients take the form $A$ = -4.447 $\pm$ 0.781, $B$ = -1.084 $\pm$ 0.089, and $C$ = -0.020 $\pm$ 0.002, where 12+log(O/H) = $A$~+~$Bx$~+~$Cx^{2}$ and $x$~=~$M_{B}$.}
We find a marginally worse RMS scatter in the vertical (abundance) axis of $\sigma$ = 0.282, an indication that our data does not support the necessity of higher-order fitting.
\\
\indent Many metallicity relation studies employing large datasets utilize binned galaxy data in effort to reduce systematic uncertainties.
For the sake of comparison, we developed a linear bisector \LZ\ relation for binned KISS galaxy data, finding no appreciable difference in the resulting fit.
Consequently, we have opted to proceed using the \LZ\ fit developed for unbinned KISS data.
Furthermore, in addition to the linear \LZ\ relation fit shown in Figure \ref{fig:LZ}, which uses only O3N2-method abundances, we have produced a linear \LZ\ relation using ``best available" metallicities.
This alternative relation fit hierarchically selected KISS galaxy abundances, using first those calculated using the \Te-method when present (42 galaxies), then subsequently abundances derived from the McGaugh model grid (680 galaxies), and finally O3N2-method abundances for the remainder (746 galaxies).
We found no appreciable improvement using this ``best available" fit, and therefore choose to maintain the O3N2-only \LZ\ relation.
This result, however, supports the reliability of our O3N2-method calibration as a means to estimate metallicities.

\subsection{Mass-Metallicity ($M_{*}$--$Z$)} 

\begin{figure} 
\includegraphics[height=8.6cm,width=8.6cm]{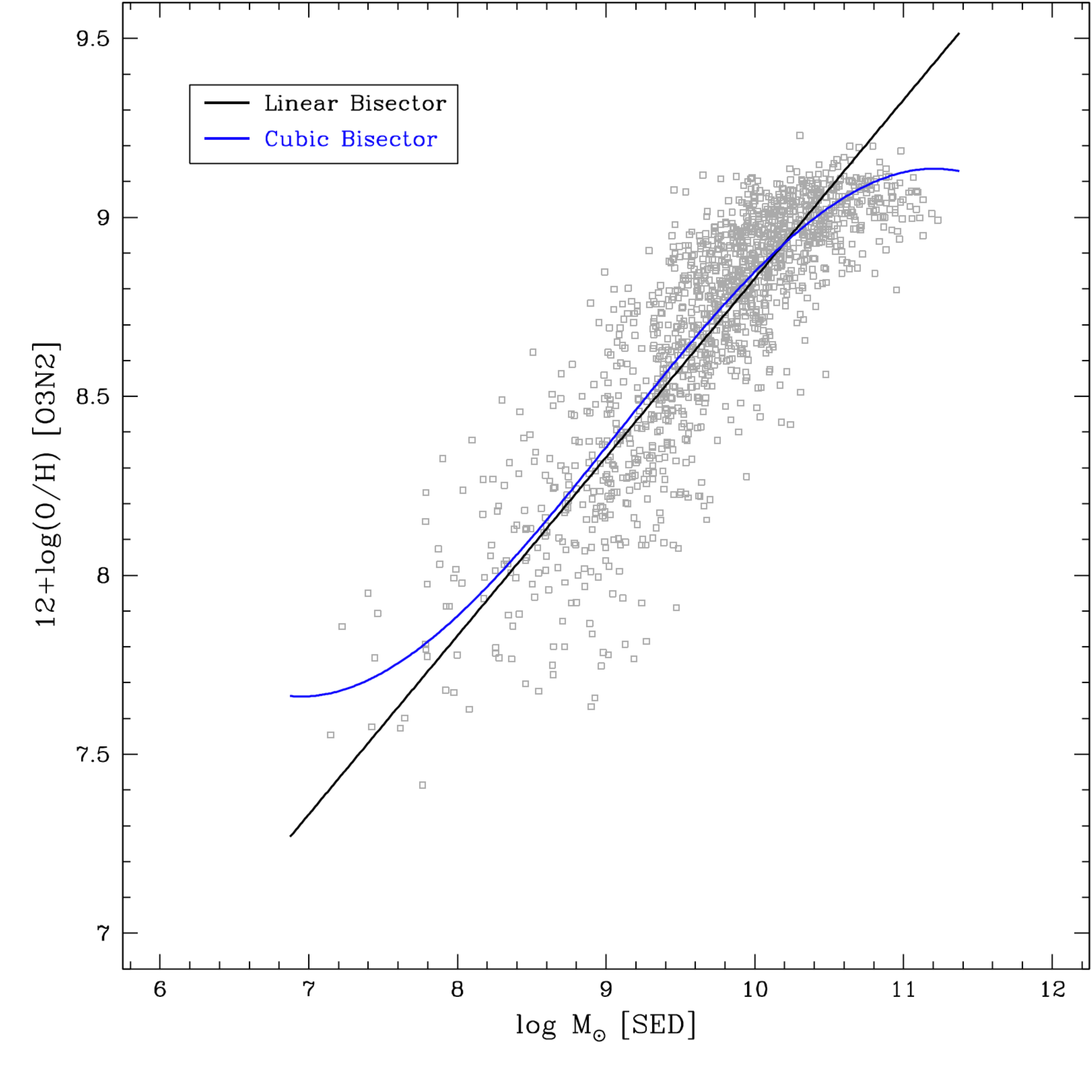}
\caption{Linear (black line) and cubic (blue line) mass-metallicity (\MZ) relation fits to the KISS sample of star-forming galaxies.
Oxygen abundances for this sample are derived via the O3N2 abundance method, while stellar masses are determined from SED fits to a variety of photometric data sets.
The linear and cubic fits are roughly coincident over the range of log $M_{*}$ where the majority of galaxy data lie.
}
\label{fig:MZ}
\end{figure}

\indent With the inclusion of new SED-fit masses for most galaxies in the KISS database (see \S2.3 and the appendices), the development of an \MZ\ relation using the self-consistent O3N2 metallicities for this sample of star-forming systems became possible for the first time (Figure \ref{fig:MZ}).
Using a method similar to that described in \S4.1 concerning the \LZ\ relation, we found a functional form for this linear bisector fit of:
\[
12+\log(\mathrm{O/H}) = 3.838(\pm0.102) + 0.499(\pm0.007) \times (\log M_{*}),
\]
where $M_{*}$ is the stellar mass of the system in units of solar mass.
In total, 1,450 KISS SFGs are included in this plot.
There are 18 fewer galaxies in the \MZ\ plot compared with the \LZ\ plot because in these few cases, the stellar mass estimates were deemed unreliable and were therefore rejected.
The RMS scatter for this relation is $\sigma$ = 0.182.
A tighter fit as compared to the \LZ\ relation (Figure \ref{fig:LZ}) is a strong indication of the reliability of the SED fitting methods employed by this study.
Scatter in luminosity due to the effect of both starbursts and varied levels of internal absorption appears to be largely mitigated by the SED fitting method.
\\
\indent Higher-order polynomials were developed in order to address abundance flattening for our KISS \MZ\ distribution.
The data seen in Figure \ref{fig:MZ}, however, appears to flatten both at high \emph{and} low metallicities.
A cubic bisector function was therefore developed to fit the data, possessing a noticeably reduced RMS scatter in the vertical (abundance) axis of $\sigma$ = 0.169.\footnote{\footnotesize Cubic bisector fit polynomial coefficients take the form $A$ = 32.307 $\pm$ 0.058, $B$ = -8.940 $\pm$ 0.020, $C$ = 1.042 $\pm$ 0.002, and $D$ = -0.038 $\pm$ 0.00008, where 12+log(O/H) = $A$~+~$Bx$~+~$Cx^{2}$~+~$Dx^{3}$ and $x$~=~log~$M_{*}$.}
While many previous \MZ\ studies mandate a non-linear function to account for oxygen abundance flattening at high stellar masses, none address an equivalent flattening at low stellar masses.
A discussion of flattening at both ends of the mass distribution is presented in \S5.2.
\\
\indent We have additionally developed a linear bisector \MZ\ relation utilizing binned KISS galaxy data in a manner similar to that described in \S4.1 for comparison with studies found in the literature.
Again, we have found no appreciable difference and have opted to proceed using the linear \MZ\ relation developed for unbinned KISS data.
Finally, we have made a ``best available" KISS \MZ\ relation employing an identical algorithm for selecting the appropriate abundance for each galaxy as that described in \S4.1.
Once again, we find no appreciable improvement over the O3N2-only \MZ\ relation.

\subsection{Fundamental Metallicity Relation (FMR)} 

\begin{figure} 
\includegraphics[height=8.6cm,width=8.6cm]{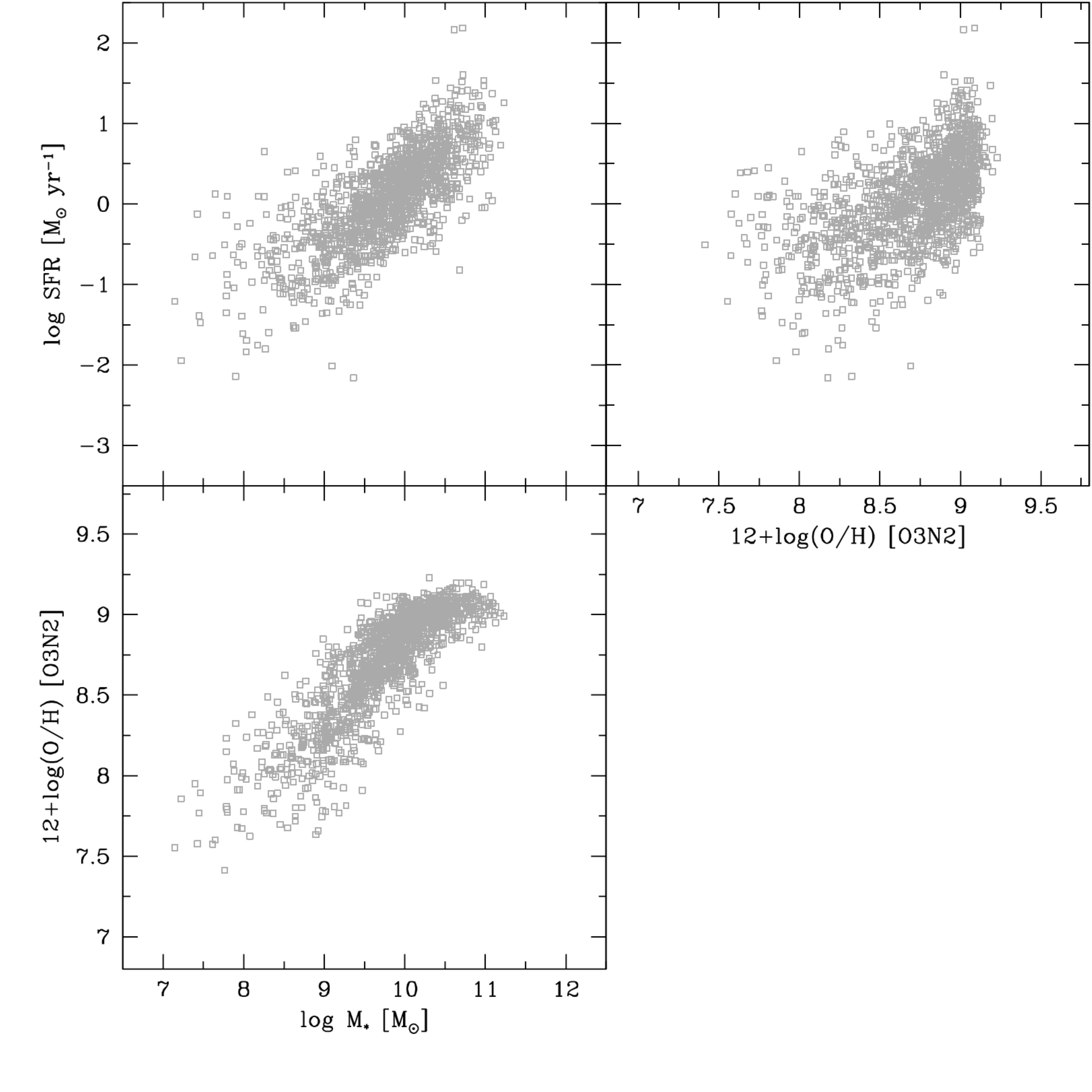}
\caption{Plots demonstrating the relationships between stellar mass and metallicity (\MZ; bottom-left), stellar mass and star formation rate ($M_{*}$--SFR; upper-left), and metallicity and star formation rate ($Z$--SFR; upper-right) for star-forming galaxies in the KISS database.
The individual correlations that exist between each of these three parameters implies the manifestation of a fundamental plane (FP) relationship, whose axes reduce to these three more familiar relations.
}
\label{fig:3par}
\end{figure}

\indent In addition to the \MZ\ relationship concerning stellar mass and gas-phase metallicity of SFGs, the so-called ``star formation main sequence" compares $M_{*}$ with SFR.
This relation is approximately linear, persists at both low- \citep{bib:Brinchmann2004, bib:Salim2007} and high-redshift \citep{bib:Daddi2007}, and is observed with SF indicators in optical \citep{bib:Tasca2014}, infrared \citep{bib:Elbaz2011}, and radio \citep{bib:Karim2011} data.
Like \MZ, the \MSFR\ relation exhibits relatively small scatter ($\sigma$ $\simeq$ 0.2 dex; \citealp{bib:Speagle2014}) and the SFR at a given value of $M_{*}$ appears to increase with redshift from $z$~=~0 to $z$~=~2, when the global star formation rate density peaks.
Recent efforts have shown that the \MZ\ and \MSFR\ relations can be seen as two-dimensional projections of a more fundamental, three-dimensional FP relating stellar mass, gas-phase metallicity, and star-formation rate.
A recognition of the dependence of $M_{*}$ and $Z$ with SFR by the study of \citet{bib:Ellison2008} represented the first step in exploring this phenomenon:
For galaxies of a given stellar mass, those with higher rates of star-formation were found to be under-enriched as compared to systems with lower star-formation rates.
A three-parameter fit to these quantities has been shown to yield an FP with smaller scatter than the \MZ\ relation, an indication that the latter exists as a projection of a more fundamental relationship.
\\
\indent \citet{bib:Lara-Lopez2010} related $M_{*}$, $Z$, and SFR using data from SDSS-DR7, finding that the \MZ\ and $M_{*}$--SFR relations are distinct circumstances of a more general relationship.
Simultaneously, the study of \citet{bib:Mannucci2010} produced a similar three-dimensional relationship between these same parameters, generating a surface that they referred to as the Fundamental Metallicity Relation (FMR).
In this formulation, metallicity $Z$ is compared with a term $\mu$ represented by a combination of $M_{*}$ and SFR, where $\mu$~$\equiv$~log($M_{*}$)~--~$\alpha$~log(SFR).
\\
\indent The three plots of Figure \ref{fig:3par} illustrate the correlations between $M_{*}$ and $Z$ (bottom-left), $M_{*}$ and SFR (upper-left), and $Z$ and SFR (upper-right) for star-forming galaxies in the KISS database.
SFR was determined for KISS galaxies by measurement of \HA\ luminosity through the relations in \citet{bib:Kennicutt1998} with a \citet{bib:Salpeter1955} IMF:
\[
\mathrm{SFR} = \frac{L_{\mathrm{H\alpha}}}{7.9 \times 10^{42}},
\]
where SFR is in M$_{\odot}$ yr$^{-1}$ and $L_{\mathrm{H\alpha}}$ is the \HA\ luminosity in erg s$^{-1}$.
\HA\ luminosities were derived from the original KISS objective-prism observations, which encompass emission coming from the entire galaxy.
These objective-prism spectra provide a more comprehensive measure of the \HA\ flux for each galaxy than is possible with long-slit or fiber spectra.
\\
\indent The individual correlations found in Figure \ref{fig:3par} imply the manifestation of an FP relationship, which projects to the \MZ, $M_{*}$--SFR, and $Z$--SFR relations.
Finding the value of $\alpha$ that minimizes the scatter in the $\mu$~$\equiv$~log($M_{*}$)~--~$\alpha$~log(SFR) relationship should produce a tighter fit than any of the other three relations alone.
In order to compare the KISS galaxy sample to other studies found in the literature, we have developed an FMR utilizing self-consistently derived values of $M_{*}$, $Z$, and SFR.
\begin{figure} 
\includegraphics[height=8.6cm,width=8.6cm]{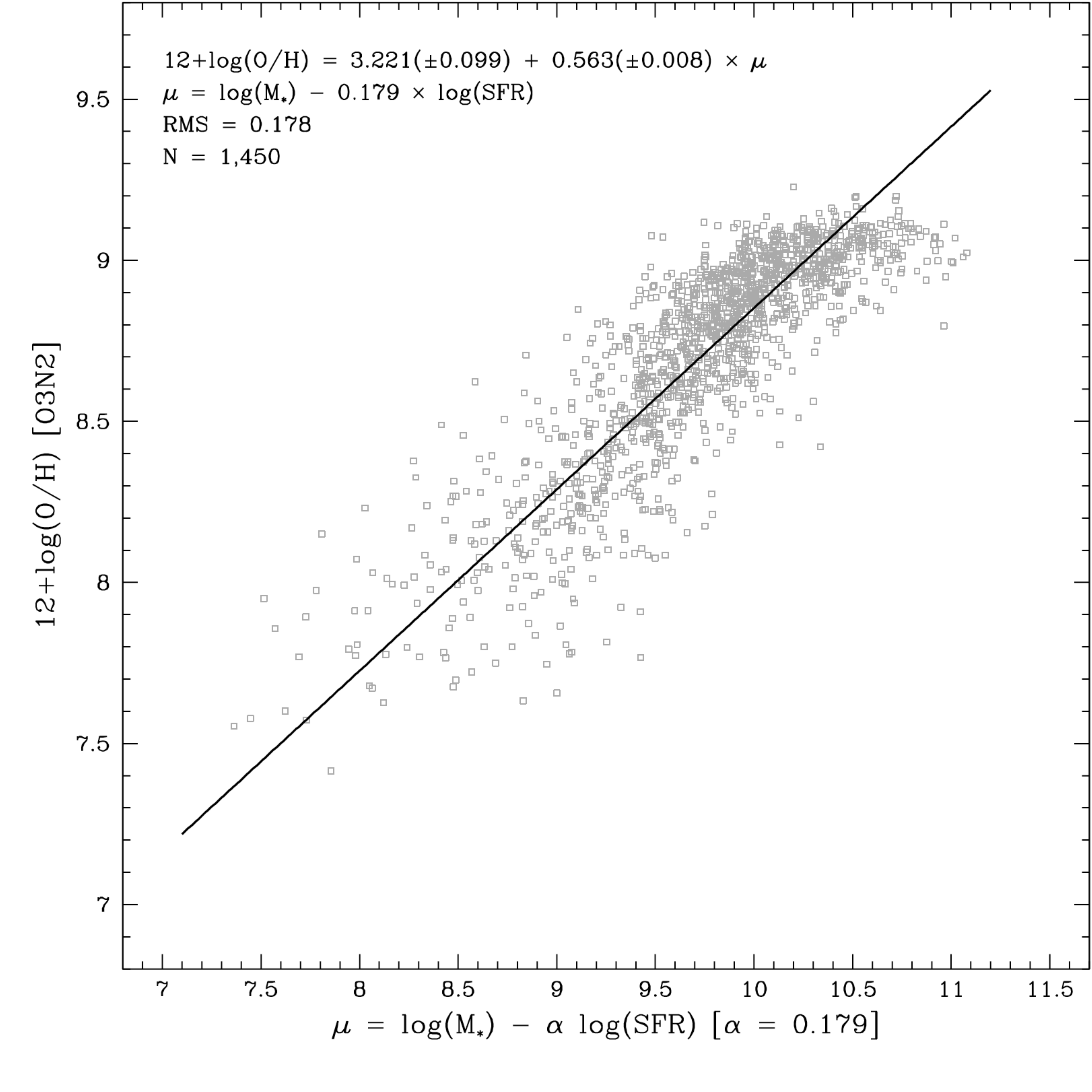}
\caption{KISS database FMR with minimized uncertainty about the vertical (abundance) axis.
The ``best" value of $\alpha$ was found to be $\alpha$ = 0.179.
The associated RMS scatter for this minimized value of $\alpha$ is 0.178.
}
\label{fig:FMR_minmu}
\end{figure}
Minimum vertical scatter is indicative of the ``best" combination of the $M_{*}$ and SFR terms in order to determine galaxy metallicity with the lowest uncertainty.
The $\alpha$ term is varied between 0 and 1, such that for $\alpha$ = 0, $\mu$ corresponds with log($M_{*}$), reducing the FMR to the \MZ\ relation (e.g., the lower-left panel of Figure \ref{fig:3par}).
Alternatively, when $\alpha$ = 1, $\mu$ becomes log(1/sSFR) = --log(sSFR), where sSFR is the specific star-formation rate (SFR normalized by stellar mass; \citealp{bib:Mannucci2010}).
Minimum scatter, then, is found for some intermediate value of $\alpha$.
\\
\indent In order to determine the best value of $\alpha$ to minimize scatter about $\mu$ in the FMR, RMS uncertainties about values of $\alpha$ were calculated for the KISS sample of star-forming galaxies.
For our sample, we have found the value of $\alpha$ that minimizes scatter in the FMR to be $\alpha$ = 0.179.
The accompanying oxygen abundance versus $\mu$ plot and linear bisector fit to the data is shown as Figure \ref{fig:FMR_minmu}, taking the form:
\[
12+\log(\mathrm{O/H}) = 3.221(\pm0.099) + 0.563(\pm0.008) \times\ \mu,
\]
with RMS scatter in the abundance (vertical) direction of $\sigma$ = 0.178.
Discussion of our derived FMR, including the value of minimized $\alpha$, and a comparison to previous work is deferred to the next section.

\section{Discussion} 

\indent Many extant studies have produced fits for metallicity relations utilizing a wide variety of sample selection techniques and abundance calibrations.
The adopted choices for these quantities have profound ramifications on the resultant forms of such metallicity relations.
Notably, \citet{bib:KewleyEllison2008} demonstrated the effects an application of ten different SEL abundance determination methods on a set of SFGs has on the development of the \MZ\ relation.
For a given stellar mass, the adopted choice of abundance calibration can severely affect the slope and alter the estimated metallicity by up to $\sim$0.7 dex (see their Figure 2).
Clearly, this effect has significant impact on the resulting functional form of the \LZ, \MZ, and $M_{*}$--$Z$--SFR relations for any given sample of systems.
Careful consideration must be given to such issues in order to provide proper observational constraint for theoretical models.
\\
\indent With the establishment of new metallicity relationships utilizing the updated KISS sample of SFGs, we are capable of making direct comparisons with previous studies available in the literature.
We investigate how sample selection and metallicity estimation method affect galaxian metallicity relations.
Differences are subsequently manifest as variations in the results afforded by models of chemical evolution.
In the following subsections, we explore the differences in \LZ, \MZ, and the FMR between our KISS-derived relations and equivalent fits prominent in the literature.
Given the large number of previous studies, we limit our comparison to a modest number of representative relations.
Furthermore, we confine our comparison to the $B$-band \LZ\ relations only, except for our discussion comparing this work with previous KISS optical and NIR \LZ\ relation fits (\citealp{bib:Salzer2005a}; see \S5.1.3).
\\
\indent In comparing the functional forms of our \LZ\ and \MZ\ relations with those available in the literature, we present the relevant coefficients, errors on the coefficients, and RMS uncertainties about the fits in Tables \ref{tab:KISSLZ} and \ref{tab:KISSMZ}, respectively.
\LZ\ relation fits summarized in Table \ref{tab:KISSLZ} are presented in the form \abun\ = $A$~+~$Bx$, where $x$~=~$M_{B}$.
\MZ\ relation fits summarized in Table \ref{tab:KISSMZ} are presented in the form \abun\ = $A$~+~$Bx$~+~$Cx^{2}$, where $x$~=~log $M_{*}$.
Functional form comparisons of optical and NIR \LZ\ relation fits developed using the KISS sample of SFGs are summarized in Table \ref{tab:LZNIR} and are presented in the form \abun\ = $A$~+~$Bx$, where $x$~=~$M_{H}$ or $M_{K}$.
Again, we note that the decision to limit our literature comparisons to linear \LZ\ relations and either linear or quadratic \MZ\ relations should not be taken to mean that we feel that these are the only viable functional forms of these relations.
Rather, these limitations were implemented to simplify our comparisons while reflecting common practices in previous studies.

\subsection{$L$--$Z$ Relation Comparison} 

\indent The \LZ\ relation fit developed by this study was constructed using a dataset of SFGs covering a wide dynamic range of luminosities and metallicities.
We have analyzed our results against other examples available in the literature in order to understand how our dataset and functional fit compares and contrasts with those of other studies.
Samples which cover a similarly large range in galaxy luminosity produce \LZ\ fits that generally parallel our fit.
Variations in metallicity calibration, however, can create substantial differences in the degree to which the fits overlap the current dataset.
Many works focusing on obtaining \Te-method oxygen abundances for smaller sets of ELGs arrive at substantially different forms of \LZ.
We address each variety of galaxy sample and their respective \LZ\ fits separately and compare them to our results presented in \S4.1.
In addition, we compare our $B$-band \LZ\ relation fit with KISS \LZ\ relations developed for the $H$- and $K$-bands, to investigate the effect of the expanded KISS metallicity sample of galaxies since the previous study of \citet{bib:Salzer2005a}.

\subsubsection{Large Sample $L$--$Z$ Relations with SEL-Method Metallicities} 

\begin{figure*} 
\plotone{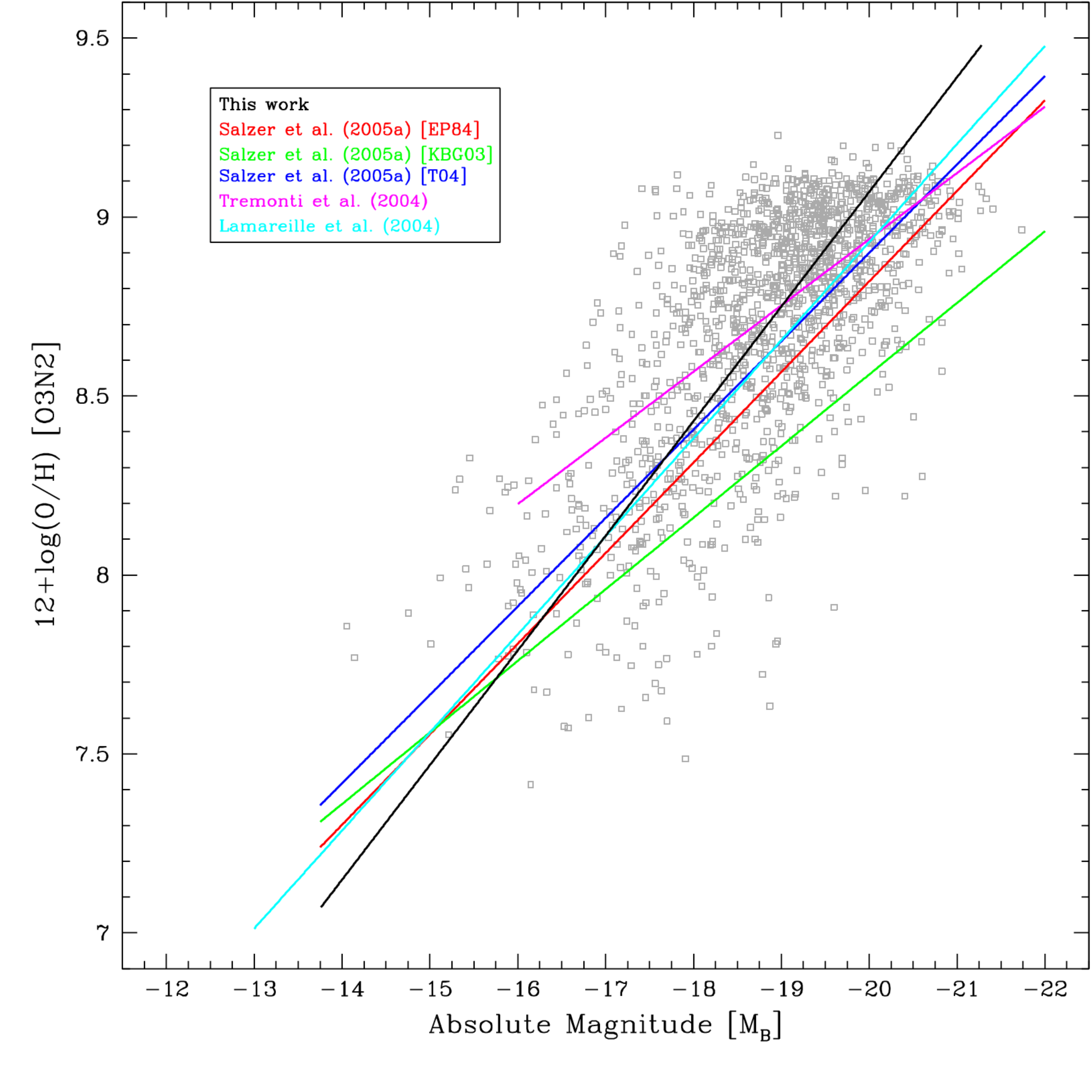}
\caption{\footnotesize A comparison of \LZ\ relation fits made from the KISS database of SFGs and other studies utilizing a wide range of galaxy luminosities.
The black line represents the current study, while the red, green, and blue lines are from \citet{bib:Salzer2005a} utilizing different metallicity estimation calibrations:
Edmunds \& Pagel (1984; EP84), Kennicutt et al.\ (2003; KBG03), and Tremonti et al.\ (2004; T04), respectively.
The magenta line is from \citet{bib:Tremonti2004}, with abundances computed via simultaneous fits to various important strong nebular emission lines.
The cyan line is from \citet{bib:Lamareille2004}, with abundances computed via the \citet{bib:McGaugh1991} SEL calibration.
The current work demonstrates a steeper slope and higher intercept than previous KISS studies.
Gray points are KISS SFGs.
}
\label{fig:KISS_LZ_compare}
\end{figure*}

\indent Obtaining oxygen abundances for a large ($\textgreater$ $\sim$1,000) sample of SFGs is an observationally challenging project that requires huge amounts of telescope time and data processing power.
Projects such as the SDSS have provided spectral data for multitudes of star-forming systems, however the observational depth obtained for a given galaxy is rarely sufficient to provide for measurable flux in the emission lines of temperature-sensitive species such as \OIII$\lambda$4363.
Possessing direct-method metallicities for even a small fraction of a large-scale sample of SFGs is therefore rare, and direct-method abundances are nearly always limited to lower-luminosity systems.
Studies which employ spectral stacking (e.g., \citealp{bib:AndrewsMartini2013, bib:Curti2017}) attempt to circumvent this problem, but may be subject to substantial  biases (see \S5.3).
Empirical strong-line methods are therefore employed to estimate the oxygen abundances of such large galaxy samples.
\\
\indent Utilizing KISS photometry and the initial calibration of the Coarse abundance method to measure galactic metallicity, \citet{bib:MelbourneSalzer2002} calculated an \LZ\ relation fit for the KISS sample of star-forming galaxies with a slope of -0.267 and with an RMS scatter of $\sigma$ = 0.27.
This early KISS-based study included only 519 galaxies with suitable spectral data.
The followup study of \citet{bib:Salzer2005a} included more objects with derived abundances (765), and employed multiple \LZ\ functional fits developed from three separate sets of KISS galaxy metallicity values.
\citet{bib:Salzer2005a} utilized three distinct variations of the SEL \R\ abundance indicator to provide calibration points for their Coarse-method metallicity relations, including Edmunds \& Pagel (1984; EP84), Kennicutt et al.\ (2003; KBG03), and Tremonti et al.\ (2004; T04).
The functional coefficients and associated RMS scatters for these fits are found in Table \ref{tab:KISSLZ}.
We refer the reader to \citet{bib:Salzer2005a} \S\S 3 \& 4 for more details.
\\
\indent Following our efforts to supplement and update the KISS database as described in \S2.2, we have produced an updated KISS \LZ\ relation.
Compared to previous KISS studies, our new fit is now marginally steeper and possesses a slightly lower intercept (Figure \ref{fig:KISS_LZ_compare}).
The \LZ\ fits from \citet{bib:Salzer2005a} developed using three separate SEL-method parameterizations demonstrate the spread in metallicity that can arise depending on the choice of calibration:
At the high-luminosity end, for example, the KBG03 method produces a metallicity $\sim$0.4 dex lower than that of the T04 method for the same sample of galaxies.
Our fit to the updated KISS sample with the new O3N2-method abundances remains generally consistent with previous forms of the relation derived for galaxies in the KISS database.
\\
\indent The seminal study undertaken by \citet{bib:Tremonti2004} produced an \LZ\ relation using 53,400 ELGs from SDSS.
The \citet{bib:Tremonti2004} relation exhibits a shallower slope as compared to our work and the previous KISS studies (Figure \ref{fig:KISS_LZ_compare}).
An explanation for the comparatively shallow slope lies in the SEL metallicity estimation method employed, which flattens somewhat at high luminosities (see \citealp{bib:Tremonti2004} \S3.1 for details).
Furthermore, the magnitude-limited nature of the \citet{bib:Tremonti2004} sample means that the resulting \LZ\ fit is heavily weighted to the higher luminosity galaxies ($M_{B}$ $\textless$ -20), whereas the KISS sample has a more uniform luminosity distribution.
A linear fit made to this sample in only the luminosity range leading up to the region where the metallicity distribution starts to flatten produces a slope more consistent with that of the current study.
\\
\indent Utilizing spectral data from the 2dF Galaxy Redshift Survey (2dFGRS) of 6,387 SFGs, the study of \citet{bib:Lamareille2004} has also produced an \LZ\ relation fit.
Because it utilizes the \citet{bib:McGaugh1991} SEL-method calibration for estimating metallicities, this study makes an ideal comparison sample for the current KISS-based study.
As seen in Figure \ref{fig:KISS_LZ_compare}, the \LZ\ fit of \citet{bib:Lamareille2004} closely matches that of the current study, a consequence of representative sample selection and choice of SEL-method metallicity calibration.
\\
\indent The choice of metallicity calibration for large samples of SFGs has profound effects on the form of the \LZ\ relation.
Nonetheless, it is clear that slope values in the range of $\sim$~0.2--0.3 are favored.
The polynomial coefficients for the functional forms of these fits are summarized in Table \ref{tab:KISSLZ}.

\begin{deluxetable*}{cccc} 
\tablewidth{0pt}
\tablecaption{Luminosity-Metallicity Relation Polynomial Coefficients for Functional Forms}
\tablehead{\colhead{\LZ\ Study}&\colhead{$A$}&\colhead{$B$}&\colhead{RMS}}
\startdata
\noindent SEL-Method Metallicities																\\
\hline
\noindent This work						&	2.664 $\pm$ 0.170	&	-0.320 $\pm$ 0.009	&	0.280	\\
\noindent \citet{bib:MelbourneSalzer2002}		&	3.60 $\pm$ 0.20	&	-0.267 $\pm$ 0.009	&	0.27		\\
\noindent \citet{bib:Salzer2005a} [EP84]		&	3.32 $\pm$ 0.06	&	-0.280 $\pm$ 0.003	&	0.292	\\
\noindent \citet{bib:Salzer2005a} [KBG03]		&	4.18 $\pm$ 0.06	&	-0.222 $\pm$ 0.003	&	0.245	\\
\noindent \citet{bib:Salzer2005a} [T04]		&	3.57 $\pm$ 0.06	&	-0.271 $\pm$ 0.003	&	0.277	\\
\noindent \citet{bib:Tremonti2004}			&	5.238 $\pm$ 0.018	&	-0.185 $\pm$ 0.001	&	0.1		\\
\noindent \citet{bib:Lamareille2004}			&	3.45	 $\pm$ 0.09	&	-0.274 $\pm$ 0.005	&	0.27		\\
\hline																							
\noindent Direct-Method Metallicities																\\
\hline																							
\noindent \citet{bib:Skillman1989}			&	5.50				&	-0.153			&	0.16		\\
\noindent \citet{bib:RicherMcCall1995}		&	5.67 $\pm$ 0.48	&	-0.147 $\pm$ 0.029	&	$\ldots$	\\
\noindent \citet{bib:Guseva2009}			&	5.745 $\pm$ 0.199	&	-0.134 $\pm$ 0.012	&	$\ldots$	\\
\noindent \citet{bib:vanZeeHaynes2006}		&	5.65 $\pm$ 0.17	&	-0.149 $\pm$ 0.011	&	0.15		\\
\noindent \citet{bib:vanZee2006}			&	5.67 $\pm$ 0.21	&	-0.151 $\pm$ 0.014	&	0.21		\\
\noindent \citet{bib:Lee2006}				&	5.94 $\pm$ 0.27	&	-0.128 $\pm$ 0.017	&	0.161	\\
\noindent \citet{bib:Berg2012}				&	6.27 $\pm$ 0.21	&	-0.11 $\pm$ 0.01	&	0.15		\\
\noindent \citet{bib:Lee2004}				&	5.37 $\pm$ 0.46	&	-0.159 $\pm$ 0.029	&	0.260	\\
\noindent \citet{bib:Haurberg2015}			&	6.21 $\pm$ 0.11	&	-0.113 $\pm$ 0.007	&	$\ldots$	\\
\noindent This work [\Te\ only]				&	6.543 $\pm$ 0.116	&	-0.084 $\pm$ 0.007	&	0.216	\\
\enddata
\label{tab:KISSLZ}
\tablecomments{Polynomial coefficients are presented in the form of \abun~=~$A$~+~$Bx$, where $x$~=~$M_{B}$.}
\end{deluxetable*}

\subsubsection{$L$--$Z$ Relations with Direct-Method Metallicities} 

\begin{figure*} 
\plotone{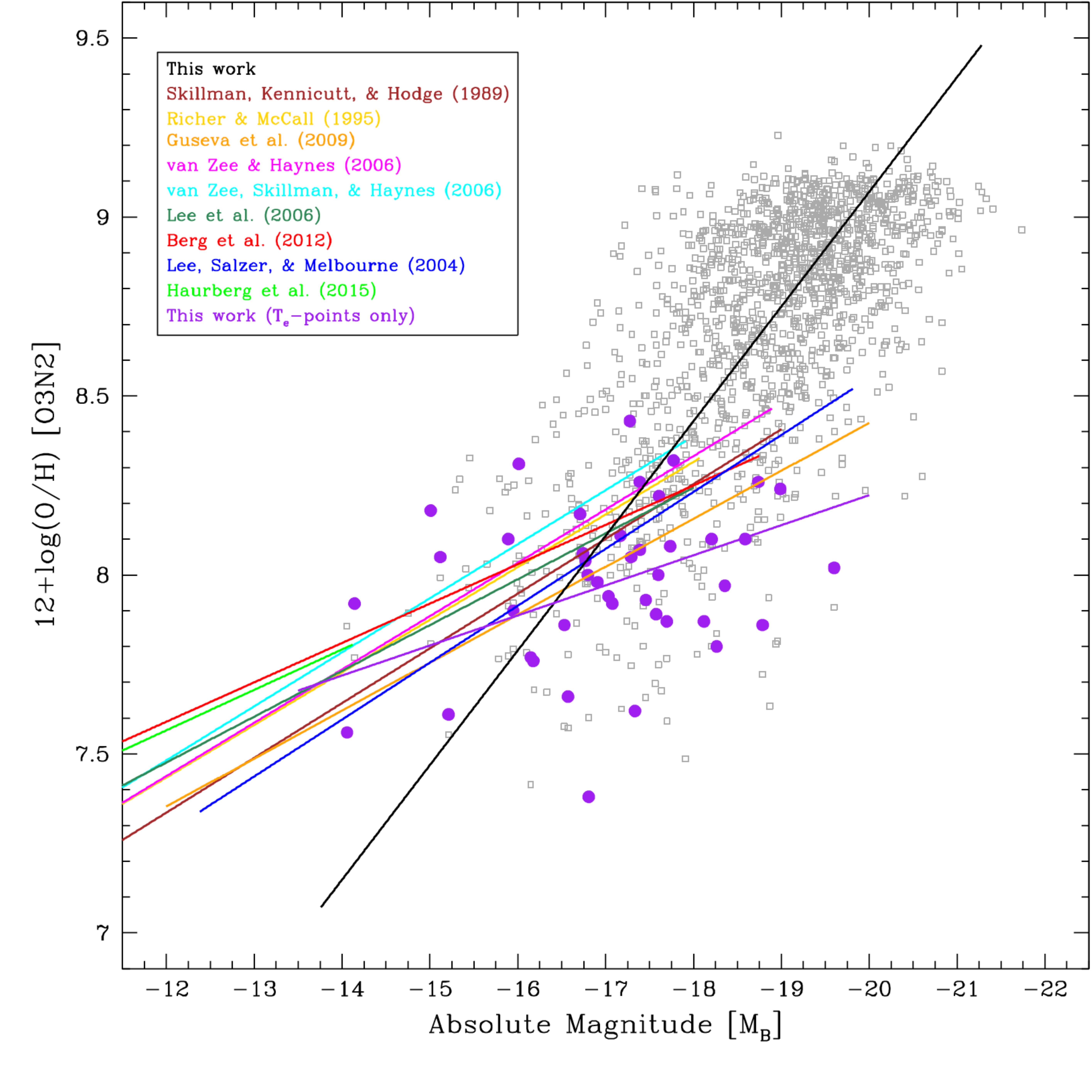}
\caption{\footnotesize A comparison of \LZ\ relation fits developed from samples of systems with \Te-method oxygen abundances with that of this study made utilizing the KISS database of SFGs.
Gray points are KISS SFGs, while purple points represent the KISS galaxies with \Te-method abundances.
The \LZ\ fits made from \Te-abundances are noticeably similar in slope, and in addition are consistently limited in range to relatively low luminosities.
The luminosity ranges plotted for each fit represent the actual ranges of the data used to derive the fits.
The relationship between galaxy luminosity and gas-phase oxygen abundance in this regime appears to be different than for studies examining a wider range of galaxy demographics.
}
\label{fig:LZ_Te_compare}
\end{figure*}

\indent The development of an \LZ\ relation fit utilizing robust, direct-method oxygen abundances has promise to accurately represent the metallicities of galaxies in a manner that improves upon the estimations afforded by any SEL technique.
As stated previously, however, acquiring accurate spectral data with \Te-quality data for a large sample of SFGs spanning the full range of luminosities is an enormous observational challenge.
These robust metallicity determinations are limited to galaxies at relatively low luminosities.
SEL abundance estimation techniques are therefore a necessity for at least some objects in datasets spanning the full range of galaxy luminosities.
Nevertheless, many studies exist in the literature that have developed \LZ\ relation fits to samples of SFGs which possess oxygen abundance values derived via the direct-method.
These works aim to characterize the astrophysical processes associated with star formation in low-luminosity systems.
The resulting functional fits are illustrated in comparison to this work in Figure \ref{fig:LZ_Te_compare}, with polynomial coefficients summarized in Table \ref{tab:KISSLZ}.
\\
\indent An examination of Figure \ref{fig:LZ_Te_compare} clearly demonstrates the similarities of the \LZ\ fits with one another from the available studies.
It is clear that relations derived for samples of SFGs with oxygen abundances computed by the direct-method are restricted to lower luminosities and all share approximately the same slope.
The small ($\sim$0.25 dex) spread in the abundance scale is almost certainly a consequence of small variations in the specific techniques used in determining \Te-method abundances.
It is also possible that some variations in the various direct-method \LZ\ fits are due in part to luminosity enhancements impacting specific samples.
For example, the \citet{bib:Lee2004} KISS sample and the \citet{bib:Guseva2009} SDSS sample are dominated by BCD galaxies.
The \LZ\ fits from these studies are the two located further to the right in Figure \ref{fig:LZ_Te_compare}.
The BCDs likely have their luminosities enhanced by 0.3-1.0 magnitudes relative to more quiescent dwarf irregulars that dominate other samples.
\\
\indent We point out an important effect that likely acts to bias the slopes of the \Te-method \LZ\ fits low.
Consider the galaxies in an \LZ\ diagram like Figure \ref{fig:LZ_Te_compare} located at intermediate luminosity ($M_{B}$ $\sim$ -17 to -19).
Any star-forming system with measurable temperature-sensitive emission lines at these luminosities will necessarily sit on the low-abundance side of the metallicity distribution at these luminosities.
The temperature-sensitive emission lines necessary for computation of direct-method oxygen abundances are typically too weak to measure for higher-metallicity systems (e.g., galaxies with \abun\ $\gtrsim$ 8.5).
ELGs at intermediate luminosities with direct-method metallicity measurements used in the development of \LZ\ relations, then, can only be those of sufficiently low abundance that \OIII$\lambda$4363 remains recoverable.
This precludes any star-forming system outside of the lowest metallicities, and therefore any resulting \LZ\ relationships using such systems will always be biased to lower slopes.
\\
%
%
\indent To further illustrate this bias, star-forming KISS galaxies with \Te-method abundances are shown as purple dots in Figure \ref{fig:LZ_Te_compare}.
These points are found to be restricted to only low and intermediate luminosities, consistent with expectation.
While direct-method abundances for galaxies at low-luminosities are spread evenly amongst the spread of metallicities estimated using our O3N2-method calibration, at intermediate-luminosities these galaxies are clearly restricted to the lower-abundance side of the distribution of points.
For galaxies with luminosities brighter than $M_{B}$ = -17.5, the direct oxygen abundances all lie \emph{beneath} the linear \LZ\ fit (black line) made to our full sample.
At intermediate absolute $B$-band magnitudes, then, galaxies with direct-method metallicities are biased low compared to the full sample of galaxies in that luminosity range.
A linear bisector fit (purple line) made for the \Te-method abundance sample alone demonstrates a considerably shallower slope as compared to the full-sample fit (slope = -0.084, RMS scatter $\sigma$ = 0.216).
This finding is consistent with the slopes determined by the studies which use direct-method abundances only. 
\\
\indent Even with this bias accounted for, the difference in the slope of \LZ\ for the sets of \Te-method samples and that of this and other works illustrated in Figure \ref{fig:KISS_LZ_compare} raises the possibility that the \LZ\ relation may not be linear at all.
That is, the slope of the \LZ\ relation appears to change significantly over the full range of luminosity covered by the data.
Low-luminosity star-forming systems appear to operate under somewhat different astrophysical mechanisms than those at high luminosities.
If sampled exclusively in the regime of lowest luminosities, the \LZ\ relation slope appears to flatten.
This may reflect the possibility that dwarf galaxies self-enrich to some nonzero abundance value early in their evolution (e.g., \citealp{bib:KunthSargent1983}).
This would lead to a ``floor" of the abundance distribution, which would naturally result in a flattening of the \LZ\ relation.
In essence, this result would indicate that SFGs cannot be arbitrarily metal-poor unless they are forming stars for the first time or they have experienced a large infall of pristine gas.
\HA\ Dots \citep{bib:Kellar2012} are recently discovered examples of emission-line sources that represent some of the lowest-luminosity examples of star-forming systems known.
Preliminary abundance analysis for these objects utilizing SEL techniques appears to produce results consistent with a flattened \LZ\ relation.
Continuing study of \HA\ Dot metallicities, including new spectral observations yielding direct-method abundances (Hirschauer et al., {\it in preparation}), promises to reveal new information on the extremely low-luminosity regime.
\\
\indent We conclude that \LZ\ (and probably \MZ) relations that exclusively use \Te\ abundances will always be biased to lower slopes and are not representative of the overall metallicity characteristics of the general population of galaxies.

\subsubsection{Optical and Near Infrared $L$--$Z$ Relations} 

\begin{figure*} 
\plotone{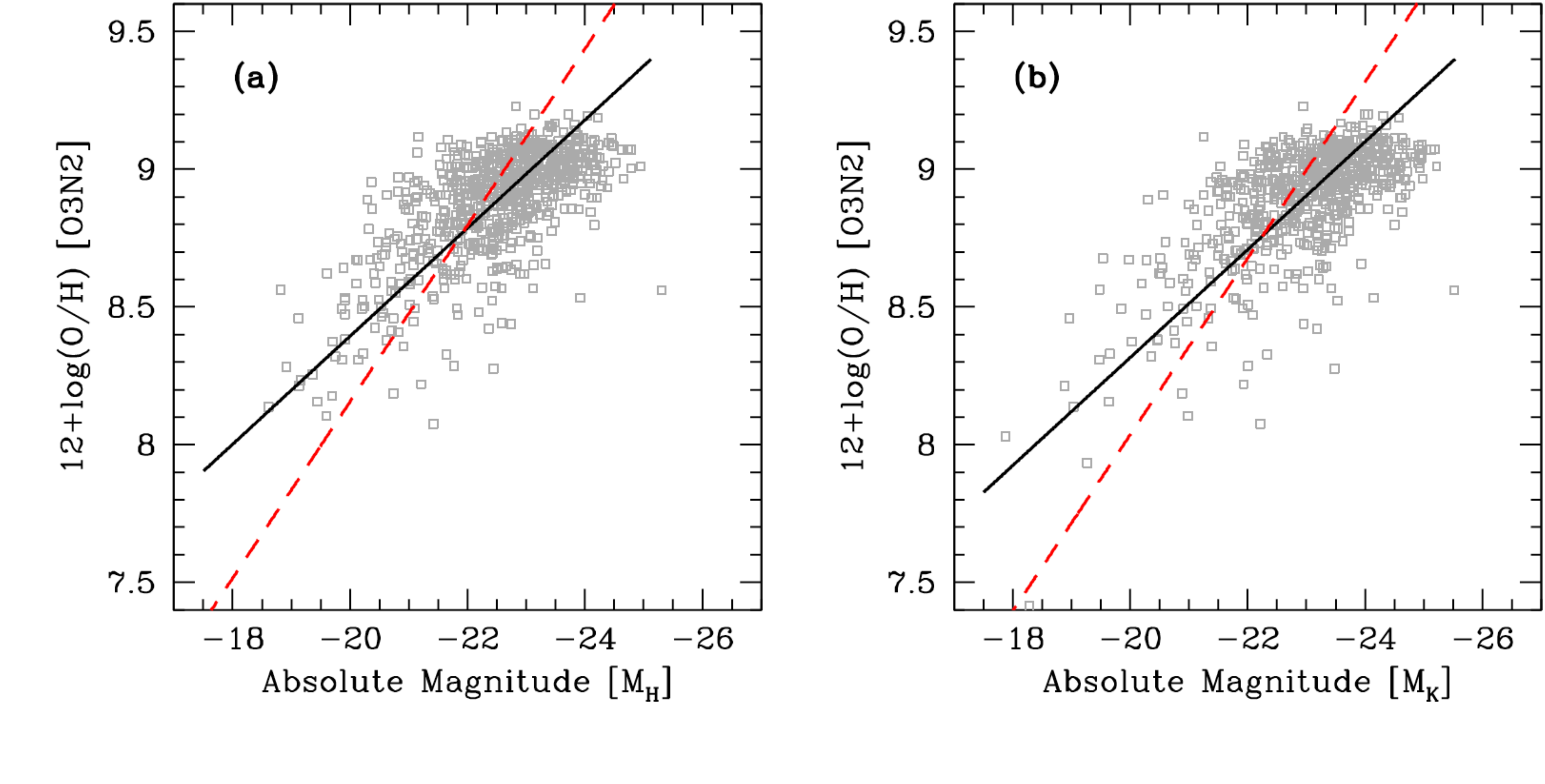}
\caption{KISS galaxy \LZ\ relations in two NIR wavelength regimes.
The left panel (a) represents $M_{H}$ data while the right panel (b) represents $M_{K}$ data.
The red dashed lines represent the slope of the $B$-band \LZ\ relation for comparison, shifted to an arbitrary $y$-intercept such that it falls roughly in the center of the plot.
Oxygen abundances ($y$-axes) are estimated using our O3N2 method defined in \S3.3.
The slope of the fits for \LZ\ relation plots becomes shallower at longer wavelengths.
}
\label{fig:4LZ}
\end{figure*}

\indent In addition to the $B$-band relations presented in \S\S5.1.1 and 5.1.2, we have made KISS galaxy \LZ\ relations measured for the $H$- and $K$-bands.
The development of metallicity relations using longer-wavelength luminosities illustrates the physical differences implicit with the measured starlight:
As the wavelength shifts from optical to NIR, the luminosity begins to trace the many lower-mass, long-lived red giant stars rather than the relatively small number of higher-mass, short-lived O- and B-type stars.
In addition, NIR wavelengths are less affected by internal absorption effects than optical light.
A progression to longer wavelengths therefore has the effect of more representatively tracing the stellar population of a given system.
\\
\indent Linear bisector \LZ\ relation fits in $H$- and $K$-bands are shown as Figure \ref{fig:4LZ}.
The left panel (a) represents $M_{H}$ data while the right panel (b) represents $M_{K}$ data.
These NIR data are from the 2MASS database; see \citet{bib:Salzer2005a} for details.
For the sake of comparison, red dashed lines representing the slope of the $B$-band \LZ\ relation, with an arbitrary vertical shift to place it roughly in the center of the plot, are included.
The shallower slope of the NIR \LZ\ relation fits as compared to that of the optical data is consistent with longer wavelengths being a more representative tracer of stellar content and lower internal absorption.
\\
\indent In comparison to the \citet{bib:Salzer2005a} study of KISS \LZ\ relations in optical and NIR, the recent addition of more spectral data of KISS galaxies has roughly doubled the number of data points in each case.
The polynomial coefficients for all fits are summarized in Table \ref{tab:LZNIR}.
While for optical data, the KISS \LZ\ relation slopes are somewhat \emph{steeper} than were found in \citet{bib:Salzer2005a}, the NIR data presents KISS \LZ\ relation slopes that are slightly \emph{shallower} than the previous study.
The slopes of -0.196 are probably close to the ``true" value, since they are derived using nearly absorption-free luminosities.
In addition, the scatter in the two NIR fits ($\sigma$ = 0.168 in $H$ and $\sigma$ = 0.178 in $K$) are slightly \emph{lower} than the scatter in our \MZ\ relation ($\sigma$ = 0.182).
This finding suggests that at NIR wavelengths, the \LZ\ relation is as reliable a tracer of stellar content in galaxies as the \MZ\ relation.
As pointed out by \citet{bib:Salzer2005a}, however, the 2MASS survey tends to undersample the intermediate- and low-luminosity galaxies, resulting in an \LZ\ relation that is not as representative of the full range of galaxies present in KISS.

\begin{deluxetable*}{ccccc} 
\tablewidth{0pt}
\tablecaption{KISS Optical and NIR Linear Luminosity-Metallicity Relation Polynomial Coefficients for Functional Forms}
\tablehead{\colhead{\LZ\ Study}&\colhead{$N$}&\colhead{A}&\colhead{B}&\colhead{RMS}}
\startdata
\noindent $H$-band																							\\
\hline
\noindent This work							&	739		&	4.469 $\pm$ 0.201	&	-0.196 $\pm$ 0.006	&	0.168	\\
\noindent \citet{bib:Salzer2005a} [EP84]			&	399		&	3.91 $\pm$ 0.09	&	-0.215 $\pm$ 0.003	&	0.225	\\
\noindent \citet{bib:Salzer2005a} [KBG03]			&	399		&	3.98 $\pm$ 0.10	&	-0.201 $\pm$ 0.004	&	0.225	\\
\noindent \citet{bib:Salzer2005a} [T04]			&	399		&	4.95 $\pm$ 0.08	&	-0.172 $\pm$ 0.003	&	0.166	\\
\hline
\noindent $K$-band																							\\
\hline
\noindent This work							&	666		&	4.398 $\pm$ 0.218	&	-0.196 $\pm$ 0.007	&	0.178	\\
\noindent \citet{bib:Salzer2005a} [EP84]			&	370		&	3.92 $\pm$ 0.09	&	-0.212 $\pm$ 0.003	&	0.235	\\
\noindent \citet{bib:Salzer2005a} [KBG03]			&	370		&	4.03 $\pm$ 0.10	&	-0.195 $\pm$ 0.004	&	0.233	\\
\noindent \citet{bib:Salzer2005a} [T04]			&	370		&	4.85 $\pm$ 0.09	&	-0.173 $\pm$ 0.003	&	0.179	\\
\enddata
\label{tab:LZNIR}
\tablecomments{Polynomial coefficients are presented in the form of \abun~=~$A$~+~$Bx$, where $x$~=~$M_{H}$ or $M_{K}$.}
\end{deluxetable*}

\subsection{$M_{*}$--$Z$ Relation Comparison} 

\begin{figure*} 
\plotone{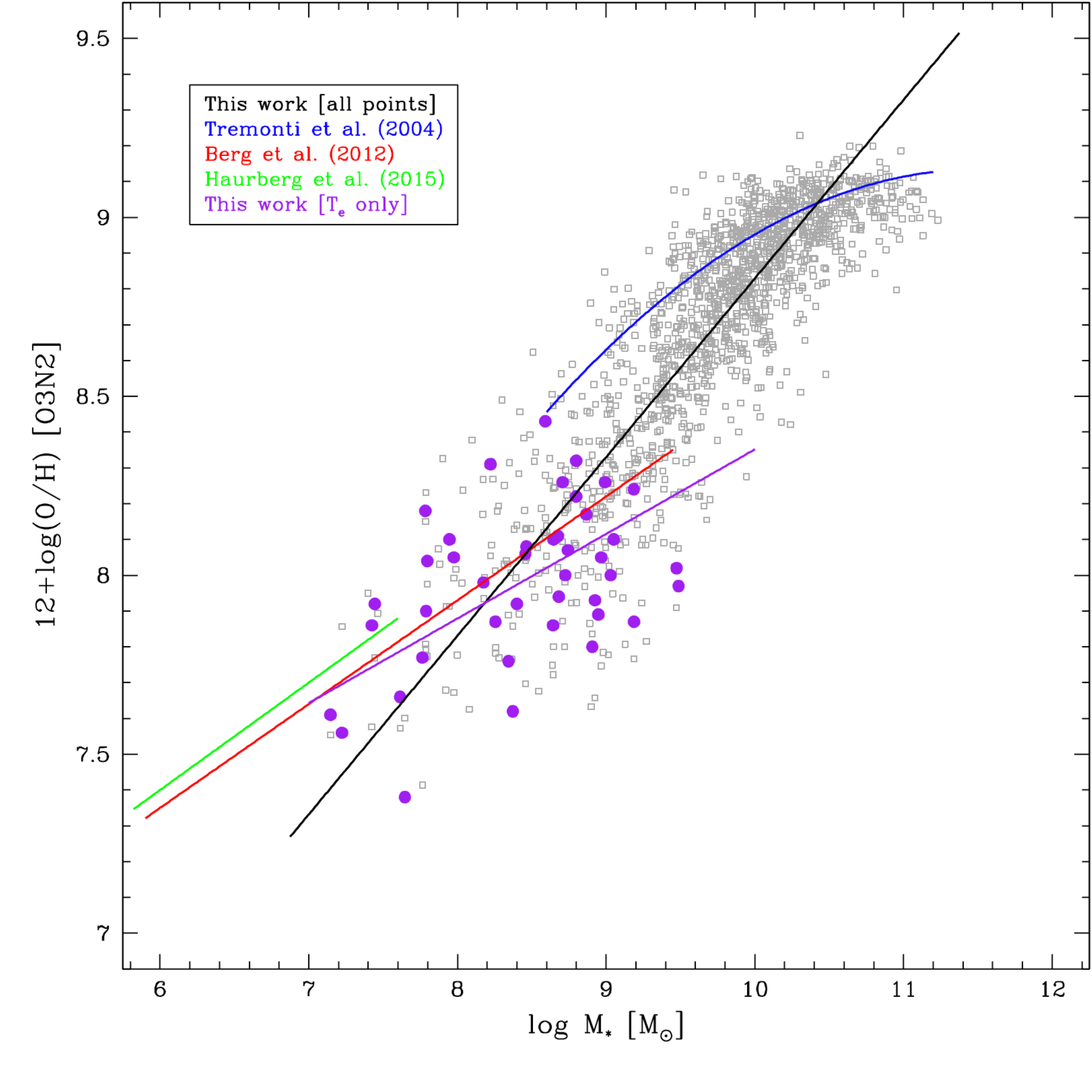}
\caption{\footnotesize Comparison of \MZ\ relations developed from this study (black line), Tremonti et al.\ (2004; blue line), Berg et al.\ (2012; red line), and Haurberg et al.\ (2015; green line).
The \citet{bib:Tremonti2004} quadratic relation demonstrates a pronounced curvature and flattening of oxygen abundance at the high-mass end.
Conversely, the \citet{bib:Berg2012} and \citet{bib:Haurberg2015} linear fits possess shallower slope, consistent with these samples being comprised solely of \Te-abundance selected galaxies representative of only low-mass, low-abundance systems.
KISS galaxies with \Te-method metallicities are plotted as purple points, illustrating a similar bias toward lower abundances at intermediate masses demonstrated with the \LZ\ plot of Figure \ref{fig:LZ_Te_compare}. 
A linear bisector fit to these points (purple line) shows yet shallower slope.
}
\label{fig:MZ_compare}
\end{figure*}

\indent As shown in \S4.2, we have produced an \MZ\ relation using the KISS database of SFGs for the first time.
Here we compare our derived result with other studies from the literature in Figure \ref{fig:MZ_compare}.
In each case, the fit is illustrated over the range of applicable masses as stated from the respective studies.
Similar to our discussion of the \LZ\ relation in \S5.1, we compare our derived fit in order to better understand the astrophysical relationship of stellar mass and gas-phase oxygen abundance.
We emphasize that the researchers' adopted choice of metallicity calibration can have a substantial effect on both the slope and abundance scale for a given set of star-forming systems \citep{bib:KewleyEllison2008}.
\\
\indent Galaxian stellar masses ($M_{*}$) for the KISS sample were computed via an SED-fitting routine (described in \S2.3) which utilizes multi-wavelength photometric data and accounts for effects such as dust absorption and star formation in a manner that adoption of a single mass-to-light ($M_{*}/L$) ratio conversion cannot.
Techniques that utilize luminosity and color to estimate mass (e.g., \citealp{bib:BrinchmannEllis2000, bib:BelldeJong2001}) can be hugely influenced for galaxies undergoing starbursts as compared to giant spirals or ellipticals, as a significant fraction of the light output is attributable to comparatively few massive, luminous stars.
The KISS SED fits are capable of tracing the older stellar populations, which constitute a far greater proportion of a galaxy's mass, as well as account for the effects of internal absorption, which can be significant.
They are therefore able to offer a more representative estimation of $M_{*}$ than $M_{*}/L$ ratios.
%
\\
\indent Some studies of the \MZ\ relation find a flattening of abundance values at high stellar mass which produces a pronounced curvature.
This necessitates the use of a higher-order fitting function.
The \citet{bib:Tremonti2004} study of $\sim$53,400 ELGs from SDSS provides a functional fit given by a quadratic equation that is reported as valid over the range of 8.5 $<$ ($\log$ $M_{*}$) $<$ 11.5.
The 1~$\sigma$ scatter of the data around the median of the distribution is $\pm$0.10 dex.
The relation shows a fairly linear appearance between 10$^{8.5}$ and 10$^{9.5}$ $M_{\odot}$.
In this mass regime the \citet{bib:Tremonti2004} relation parallels our fit, modulo a vertical offset which is attributable to differences in metallicity calibration.
Beyond 10$^{9.5}$ $M_{\odot}$, however, the \citet{bib:Tremonti2004} \MZ\ relation appears to flatten.
This is interpreted by \citet{bib:Tremonti2004} as evidence that there exists some fundamental limit to the amount of enrichment a star-forming system can achieve.
While the higher-order fit to the KISS \MZ\ data does indicate a modest flattening at higher masses (Figure \ref{fig:MZ}), the curvature is not as evident as that seen in the \citet{bib:Tremonti2004} data.
\\
\indent In addition to the large-scale \MZ\ relations like those of \citet{bib:Tremonti2004}, we also compare our work to studies utilizing smaller sample sizes and more limited ranges of galaxies.
\citet{bib:Berg2012} produced an \MZ\ relation to their ``Combined Select" sample of low-luminosity dwarf galaxies, finding a slope of 0.29 and with a reported RMS scatter of $\sigma$ = 0.15.
\citet{bib:Haurberg2015} similarly produced an \MZ\ fit to a set of extremely low-mass dwarf SFGs, deriving a relation with slope of 0.30, nearly identical to that of \citet{bib:Berg2012}.
As a direct comparison to the \citet{bib:Berg2012} and \citet{bib:Haurberg2015} studies, \Te-method abundance sources from the KISS sample are included in Figure \ref{fig:MZ_compare} as purple dots.
These metallicities are restricted to galaxies of low and intermediate masses only for the same reason as described in \S5.1.2.
\\
\indent The polynomial coefficients for the functional forms of these fits and RMS scatter are summarized in Table \ref{tab:KISSMZ}.
In comparison with the differences in \LZ\ relation slope between \Te-method abundance samples and the current work presented in Figure \ref{fig:LZ_Te_compare}, we note that the slope disparities remain present but are less severe for \MZ\ than for \LZ.
\\
\indent Finally, we note that the scatter in the new KISS \MZ\ relation is significantly lower than that in the KISS \LZ\ relation ($\sigma_{M_{*}-Z}$ = 0.182 versus $\sigma_{L-Z}$ = 0.280).
This is true despite the fact that the two samples are nearly identical and the same abundances are used for both fits.
This would seem to imply that the additional scatter in the \LZ\ relation is caused by an enhanced range of luminosity at a given mass.
Starburst events causing luminosity enhancements in each KISS galaxy are a major factor, particularly in the lower-luminosity systems.
At higher masses, dust absorption plays an increasing role in creating scatter in the observed luminosity.
Since the SED mass determinations account for the current star formation and the effects of absorption, the $M_{*}$ values used in our \MZ\ fit are not correspondingly affected.
Hence, our \MZ\ relation has much smaller scatter, and we take it to be more fundamental.
\begin{deluxetable*}{ccccc} 
\tablewidth{0pt}
\tablecaption{Mass-Metallicity Relation Polynomial Coefficients for Functional Forms}
\tablehead{\colhead{\MZ\ Study}&\colhead{$A$}&\colhead{$B$}&\colhead{$C$}&\colhead{RMS}}
\startdata
\noindent This work				&	3.838 $\pm$ 0.102	&	0.563 $\pm$ 0.007	&	---				&	0.182	\\
\noindent \citet{bib:Tremonti2004}	&	-1.492			&	1.847			&	-0.08026			&	0.10		\\
\noindent \citet{bib:Berg2012}		&	5.61 $\pm$ 0.24	&	0.29 $\pm$ 0.03	&	---				&	0.15		\\
\noindent \citet{bib:Haurberg2015}	&	5.60 $\pm$ 1.86	&	0.30 $\pm$ 0.25	&	---				&	$\ldots$	\\
\noindent This work [\Te\ only]		&	5.990 $\pm$ 0.118	&	0.236 $\pm$ 0.014	&	---				&	0.198	\\
\enddata
\label{tab:KISSMZ}
\tablecomments{Polynomial coefficients are presented in the form of \abun~=~$A$~+~$Bx$~+~$Cx^{2}$, where $x$~=~log~$M_{*}$.}
\end{deluxetable*}

\subsection{FMR Comparison} 

\indent The availability of stellar mass estimates computed via SED fits, SFRs calculated from \HA\ emission line fluxes, and self-consistent metallicity estimates of the KISS SFGs has allowed for the formulation of an $M_{*}$--$Z$--SFR FMR.
Beginning with \citet{bib:Lara-Lopez2010} reporting the development of a fundamental plane (FP) and the original FMR work of \citet{bib:Mannucci2010}, many studies have been undertaken to combine these three parameters in effort to more fully understand the foundational astrophysics.
\\
\indent Using emission-line analysis carried out by the MPA/JHU group of ELGs from SDSS-DR7 \citep{bib:York2000, bib:Abazajian2009}, the initial FMR study carried out by \citet{bib:Mannucci2010} was comprised of 141,825 SFGs at $z$ $\sim$ 0, supplemented by a set of 182 objects taken from the literature at intermediate redshift ($z$ = 0.5--2.5) and 16 galaxies observed at high redshift ($z$ = 3--4).
The minimization of the \citet{bib:Mannucci2010} FMR scatter occurs at a value of $\alpha$ = 0.32, with residual metallicity dispersion for local systems about this relationship of $\sigma$ = $\sim$0.05 dex.
Furthermore, they found no evidence for evolution up to $z$ = 2.5, an indication that unlike the \MZ\ relation, the FMR implies a tight scaling of the properties associated with the star-formation process between the high redshift and local universe.
Some evolution in the relation seen for SFGs at $z$ $\sim$ 3.3 may come as a consequence of observational effects and selection biases, but overall some other physical mechanisms are likely to become increasingly more relevant at this epoch.
\\
\indent The evolution witnessed in the \MZ\ relation, a consequence of the chemical enrichment being less complete for higher-redshift galaxies as compared to today, can be accounted for in the FMR formulation as a consequence of increased SFR at earlier cosmological epochs.
In essence, when the stellar mass and rate of star formation are combined as the single parameter $\mu$~$\equiv$~log($M_{*}$)~--~$\alpha$~log(SFR) and compared to metallicity, the interpretation for the FMR is that at a given galaxian mass an enhancement/reduction in SFR is balanced by a corresponding reduction/enhancement in gas-phase abundance.
Followup work of \citet{bib:Mannucci2011} extends the $\alpha$ = 0.32 FMR framework with an additional collection of low-mass galaxies, finding consistent results albeit with increased scatter.
Such increasing scatter toward dwarf galaxies is attributable to a large spread in their histories and current levels of star formation \citep{bib:HunterHoffman1999, bib:Hunt2005, bib:Zhao2010}.
\\
\indent The analysis undertaken by \citet{bib:Yates2012} developed an FMR for 177,071 unique ELGs from SDSS-DR7, plus an additional 43,767 SFGs from the cosmological semi-analytic model \textsc{l-galaxies}.
Oxygen abundances are computed using two separate methods, producing two different datasets (respectively, their Samples T1 and T2).
Minimizing the dispersion of the FMR for each of their samples produces optimum values of $\alpha$, where for T1 the best $\alpha$ = 0.36, very similar to the \citet{bib:Mannucci2010} result, while for T2 the minimizing value is $\alpha$ = 0.19.
The difference in optimum $\alpha$ between these two samples may be a consequence of the exclusion of some emission lines from the Bayesian analysis, which has the effect of strengthening the metallicity dependence on SFR, particularly for lower-mass galaxies.
At a given stellar mass, the effect on SFR measurement would act to change $\mu$ in a way that would cause $\alpha$ to be smaller for the T2 as compared to the T1.
These minimizing values of $\alpha$ produce dispersions of $\sigma$ = $\sim$0.06 dex for their T1 sample and $\sigma$ = $\sim$0.08 dex for their T2 sample.
\\
\indent The study of \citet{bib:Salim2014} utilized a sample taken from the SDSS spectroscopic survey \citep{bib:Strauss2002} following the prescriptions set by \citet{bib:Mannucci2010}.
Galactic parameters of stellar mass, metallicity, and SFR were adopted from the MPA/JHU catalog.
This work argues that because SFR scales with stellar mass, a SFR normalized to mass (the specific star-formation rate; sSFR) is a more physically motivated third parameter for the FMR.
Furthermore, they found that galaxy sample selection by SFR or sSFR introduces a mass selection, and therefore offsets in \MZ\ are a consequence of relative levels of sSFR at a given mass rather than absolute SFR.
They therefore tested the formulation of the FMR by comparing $Z$ with the {\it relative} sSFR, defined as the difference between a galaxy's sSFR and the sSFR typical for galaxies of that mass, in mass bins 0.5 dex wide.
By reducing these biases, this study found that metallicity dependence on sSFR is different for extreme star-forming systems versus ``normal" SFGs, producing a broken linear fit between $Z$ and sSFR.
Samples comprised of higher SFR galaxies and/or galaxies at higher-redshifts therefore do not follow the local FMR.
Consequently, reduction in scatter between their \MZ\ and FMR relation was found to be minimal, an indication that the FMR as a three-dimensional ``surface" is not particularly thin.
\\
\indent Also utilizing a large sample of 53,444 SFGs from SDSS-DR7, \citet{bib:Wu2016} developed an FMR for local systems and found minimizing values of $\alpha$ = 0.09 ($L_{\mathrm{H\alpha}}$-derived SFR) and $\alpha$ = 0.07 ($L_{\mathrm{[O~{\footnotesize III}]}}$-derived SFR).
These minimizing values of $\alpha$ produce dispersions of $\sigma$ = $\sim$0.050 dex and $\sigma$ = $\sim$0.051 dex, respectively.
This somewhat smaller value of $\alpha$ reduces the \citet{bib:Wu2016} FMR to nearly the \MZ\ relation, indicating that from their sample there appears to be comparatively very little SFR dependence.
The application of minimum luminosity thresholds for their samples (log($L_{\mathrm{H\alpha}}$)~$>$~41.0 and log($L_{\mathrm{[O~{\footnotesize III}]}}$)~$>$~39.7, respectively) may explain why their values of optimum $\alpha$ are low as compared to other studies:
Values of SFR derived from these \HA\ and \OIII\ luminosities consequently shade the overall sample toward the high side of the distribution.
Again, for a given stellar mass we expect that $\alpha$ for these samples must be comparatively smaller, even as compared to the \citet{bib:Yates2012} T2 Sample, due to the effects of SFR measurement on $\mu$ for the \citet{bib:Wu2016} formulation of the FMR.
\\
\indent \citet{bib:AndrewsMartini2013} created an FMR utilizing $\sim$200,000 SFGs taken from SDSS, measured in bins of stacked spectra in order to recover the temperature-sensitive \OIII$\lambda$4363 emission line and therefore compute direct-method metallicities.
They selected galaxies in the range 0.027 \textless\ $z$ \textless\ 0.25 to ensure the inclusion of necessary emission lines by the SDSS spectrograph.
Minimized scatter in metallicity for their study was found at $\alpha$ = 0.66, producing a scatter of $\sigma$ = 0.13.
This value of optimum $\alpha$ is substantially larger than those found by this study and by the previous works of \citet{bib:Mannucci2010}, \citet{bib:Yates2012}, and \citet{bib:Wu2016}.
They attribute this discrepancy to the difference between direct and empirical methods for estimating oxygen abundances, noting that SEL calibrations depend upon physical properties that are correlated with SFR.
By adopting seven different SEL-method metallicity calibrations available in the literature for their sample, \citet{bib:AndrewsMartini2013} find optimum values for $\alpha$ ranging between 0.12 and 0.34, consistent with this study as well as the \citet{bib:Mannucci2010}, \citet{bib:Yates2012}, and \citet{bib:Wu2016} results.
Scatter in their FMRs developed utilizing SEL-relation abundances is quoted for two such methods:
For the \citet{bib:KobulnickyKewley2004} \R\ calibration the $\sigma$ = 0.09, while for the \citet{bib:PettiniPagel2004} N2-method calibration the $\sigma$ = 0.07.
\\
\indent With a value of minimizing $\alpha$ = 0.179, our KISS FMR is consistent with studies utilizing empirical metallicity relations.
There remains a well-known disparity in calculated metallicities between direct- and SEL-method abundance estimation methods (see discussion in \citealp{bib:Hirschauer2015}) which is manifest in the widely discrepant values of optimum $\alpha$ presented in \citet{bib:AndrewsMartini2013}.
This may be attributable in part to a similar sample bias as that described in \S5.1.2.
At higher-luminosities, \Te-method oxygen abundances are limited to metal-poor galaxies with measurable temperature-sensitive emission lines.
Metal-rich systems are therefore comparatively under-represented in direct-method samples.
\\
\indent In addition, the process of spectral stacking itself may introduce biases which influence the measurement of total oxygen abundance.
When many spectra are combined into a single spectrum representing a small range of parameter space, the increased S/N ratio allows for the measurement of diagnostic nebular emission lines that would otherwise be too weak to detect.
This averaging process, however, is intrinsically weighted toward the highest-temperature systems subsumed within the given stacking bin, corresponding to the strongest emission of temperature-sensitive emission lines such as \OIII$\lambda$4363.
Because this line is so natively weak compared to, for example, \OIII$\lambda\lambda$4959,5007, any contribution of extraordinarily strong emission from an exceptionally metal-poor object within the specific stacking bin's parameters will disproportionally influence the average \Te\ too high.
\\
\indent While the value for optimum $\alpha$ developed by this work agrees with other studies employing SEL-method metallicity estimation techniques, the RMS scatter in the FMR for our sample ($\sigma$ = 0.178) is consistently and noticeably larger.
The results from these studies are summarized in Table \ref{tab:FMR}.
Importantly, the difference in sample selection may have significant influence on the form of the FMR.
Because the KISS catalogs of ELGs include a much higher proportion of extreme star-forming systems relative to the general SDSS spectroscopic sample, this may result in the KISS sample exhibiting a larger scatter.
That is, we should expect more scatter in the KISS FMR specifically because the KISS sample includes a larger fraction of outliers in SFR than does SDSS.
\\
\indent The resulting scatter determined for the KISS FMR does {\it not} demonstrate a marked improvement over the equivalent scatter of the KISS \MZ\ relation ($\sigma$ = 0.182).
Even for a population of systems offset high from typical levels of activity such as KISS, the scatter of an FMR is expected to be significantly reduced as compared to the classic \MZ\ relation.
The study of \citet{bib:Mannucci2010} determined that some high-redshift systems ($z$~$>$~2.5) were outliers to their FMR, attributing the scatter to observational effects concerning metallicity calibrations and segregation of AGN, as well as systematic effects concerning galactic evolution with increasing redshift (see their \S5.1).
Offsets from their FMR as a consequence of elevated SFR at high redshift may mirror the increased scatter found by this study demonstrated by the extreme SFGs which comprise the KISS sample.
Based on a sample of galaxies characterized by an elevated amount of star formation, our results appear to challenge the ``fundamental" nature of the FMR.
%
\begin{deluxetable}{ccc} 
\tablewidth{0pt}
\tablecaption{Optimum $\alpha$ for FMR Studies}
\tablehead{\colhead{FMR Study}&\colhead{Optimum $\alpha$}&\colhead{RMS}}
\startdata
\noindent This work															&	0.179	&	0.178		\\
\noindent \citet{bib:Mannucci2010}												&	0.32		&	$\sim$0.05	\\
\noindent \citet{bib:Yates2012} Sample T1											&	0.36		&	$\sim$0.06	\\
\noindent \citet{bib:Yates2012} Sample T2											&	0.19		&	$\sim$0.08	\\
\noindent \citet{bib:Wu2016} $L_{\mathrm{H\alpha}}$-derived SFR						&	0.09		&	$\sim$0.050	\\
\noindent \citet{bib:Wu2016} $L_{\mathrm{[O~{\footnotesize III}]}}$-derived SFR			&	0.07		&	$\sim$0.051	\\
\noindent \citet{bib:AndrewsMartini2013}											&	0.66		&	0.13			\\
\enddata
\label{tab:FMR}
\end{deluxetable}

\begin{figure*} 
\plotone{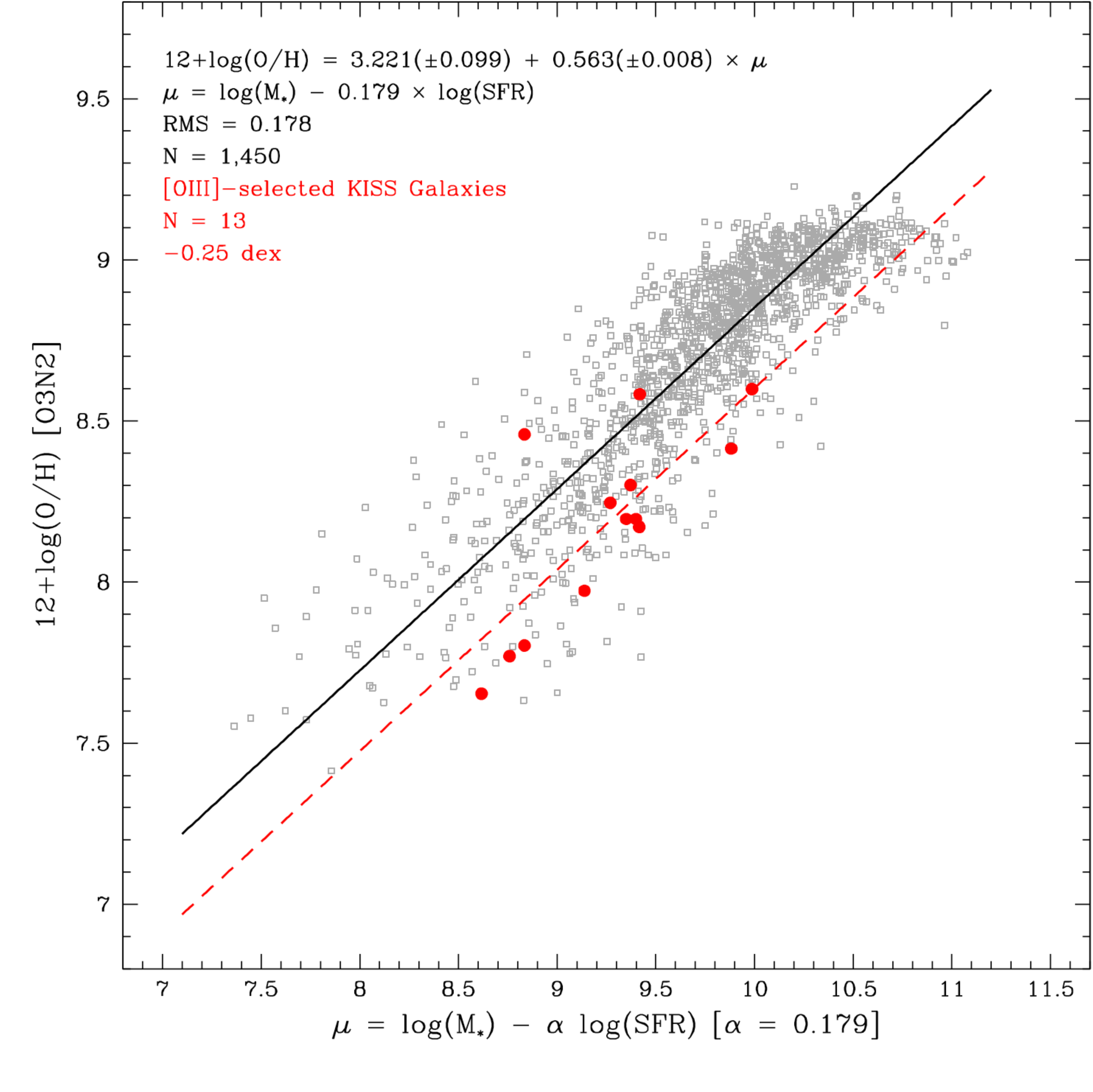}
\caption{\footnotesize KISS FMR relation with the addition of 13 high-redshift, [O~{\scriptsize III}]-selected SFGs from the KISS sample \citep{bib:Salzer2009}.
These objects demonstrate noticeable offset from the main grouping of systems, indicated by the dashed red line (offset down from the main trend line by 0.25 dex).
As a consequence of the higher SFR of KISS galaxies, we expect a larger spread as compared to samples utilizing larger numbers of more ``generic" star-forming systems.
The large offset of the extreme, high-$z$ galaxies compared to local systems calls into question the ``fundamental" nature of the FMR.
}
\label{fig:27NovFMRHZ}
\end{figure*}

\indent To further exemplify the impact of extreme star-forming systems on the FMR, we have included a set of 13 higher-redshift KISS ELGs to our FMR, shown as red dots in Figure \ref{fig:27NovFMRHZ}.
These SFGs were detected by emission of \OIII$\lambda$5007 falling into the KISS red (\HA) filter, placing them in the redshift range of $z$ = 0.29-0.42.
For more details on the characteristics of these objects, see \citet{bib:Salzer2009}.
In comparison to the main grouping of KISS galaxies, these \OIII-selected SFGs demonstrate a noticeable offset to lower oxygen abundance and/or higher $\mu$.
A red dashed line illustrates this difference as a 0.25 dex downward shift of our previously established fit, which minimizes scatter of the red dots.
This collection of systems exhibits a non-negligible offset from the already increased scatter associated with the high-activity \HA-selected KISS sample.
Line-selected galaxies detected at high redshift are fundamentally biased toward luminous systems with high rates of star-formation, similar in nature to this set of \OIII-selected SFGs.
Finding such galaxies clearly offset from the FMR may be an indication that further considerations must be made in order to truly consider this relation to be a fundamental representation of \emph{all} star-forming systems.
Our results are reminiscent of those found in \citet{bib:Salim2014}.

\section{Summary and Conclusions} 

\indent We have presented an updated metallicity analysis based on the expanded spectral database of KISS SFGs, which includes additional spectral data used for computation of metal abundances and new stellar mass estimates.
Since the most recent presentation of the KISS sample for an SEL abundance method calibration and \LZ\ relation formulation (Salzer et al.\ 2005a), new observations using the 9.2-m HET, 2.4-m MDM, and 3-m Lick telescopes have been undertaken.
These spectra have nearly doubled the number of galaxies in the database, including some with \Te-quality data for computing direct-method oxygen abundances.
The addition of \OII$\lambda$3727 emission-line data adopted from KISS galaxies matched to SDSS has allowed for the calculation of new McGaugh model grid oxygen abundances for 725 systems.
By bringing the number of robust \Te- and McGaugh-method metallicities up from 185 to 739, we have a nearly four-fold increase in objects available to calibrate an SEL-method abundance technique as compared to \citet{bib:Salzer2005a}.
Our new O3N2-method calibration has been used to provide a self-consistent metallicity estimation for all galaxies in the KISS database with suitable data.
In addition, for the first time a sophisticated SED fitting routine has supplied stellar mass estimates for KISS galaxies in a manner that mitigates the influence of luminosity enhancements from a relative few massive, luminous stars found in starburst systems, as well as self-consistently accounts for the effects of internal absorption.
\\
\indent With these estimations of oxygen abundance and stellar mass, we then developed \LZ, \MZ, and $M_{*}$--$Z$--SFR relations for the statistically representative KISS sample of ELGs.
Our linear $B$-band \LZ\ relation fit takes the form:
\[
12+\log(\mathrm{O/H}) = 2.664(\pm0.170) - 0.320(\pm0.009) \times M_{B},
\]
with an RMS scatter in abundance of $\sigma$ = 0.280.
In comparison with previous iterations of KISS \LZ, our fit is generally consistent but slightly steeper.
Differences stem from the application of our O3N2 metallicity indicator, which utilizes more high-abundance systems in its calibration than methods employed by previous KISS studies.
This \LZ\ relation fit also demonstrates a steeper slope than that of the large-scale \citet{bib:Tremonti2004} study. 
The \citet{bib:Lamareille2004} \LZ\ relation fit more closely resembles that of our KISS study, owing to similar employment of the \citet{bib:McGaugh1991} abundance calibration and an emphasis on representative sample selection.
\\
\indent A comparison of our \LZ\ relation to those developed for smaller samples of low-luminosity star-forming systems with oxygen abundances determined via the direct method reveals a dramatic difference:
Slopes for fits from the direct-abundance studies are consistently shallower in this lower-luminosity regime.
Comparisons specifying this variety of galaxy, however, are hard to observe due to their relatively paucity within larger samples such as KISS or SDSS.
This difference in slope is likely due in part to the biasing toward low abundances associated with \Te-method metallicities at intermediate luminosities.
We suggest the \LZ\ relations based solely on samples with \Te-based abundances will always be biased to shallower slopes due to this effect.
It is also possible that the \LZ\ relation exhibits a slope change at lower luminosities.
Low-luminosity systems may self-enrich to a nonzero abundance level early in their evolution (e.g., \citealp{bib:KunthSargent1983}), which would naturally lead to flattening of the \LZ\ relation.
Abundance studies of extremely low-luminosity systems like \HA\ Dots \citep{bib:Kellar2012} may shed light on this issue.
\\
\indent The development of stellar masses using SED fitting techniques allows us to construct an \MZ\ relation for the KISS galaxies for the first time.
Our linear \MZ\ relation fit takes the form:
\[
12+\log(\mathrm{O/H}) = 3.838(\pm0.102) + 0.499(\pm0.007) \times (\log M_{*}),
\]
with an RMS scatter in abundance of $\sigma$ = 0.182.
The reduction in scatter as compared to our \LZ\ fit offers strong support for the reliability of the mass derivation methods employed.
Our result demonstrates consistency with that of the \citet{bib:Tremonti2004} \MZ\ relation for systems of intermediate mass, however differences emerge at higher masses.
In contrast to our linear regression, the \citet{bib:Tremonti2004} study adopts a higher-order polynomial fit which flattens at high abundances.
This disparity appears to be attributable to differences in the abundance calibration chosen.
In addition, a comparison of our \MZ\ relation fit with those derived utilizing low-mass samples with \Te-method abundances reveals a difference in slope that parallels that seen with the \LZ\ relation.
Flattening of the \MZ\ relation slope at low-masses yields further support for rapid self-enrichment of these systems.
\\
\indent Finally, our development of an FMR for the KISS sample finds an optimum value of $\alpha$~=~0.179 which minimizes the scatter in the relationship between the term $\mu$~$\equiv$~log($M_{*}$)~--~$\alpha$~log(SFR) and oxygen abundance \abun.
This three-dimensional FP relationship of $M_{*}$, $Z$, and SFR produces the equation:
\[
12+\log(\mathrm{O/H}) = 3.221(\pm0.099) + 0.563(\pm0.008) \times\ \mu,
\]
with RMS scatter in abundance of $\sigma$ = 0.178.
Our minimizing value of $\alpha$ is consistent with other studies which use abundances determined via SEL methods.
These are not consistent, however, with methods utilizing stacked spectra to recover temperature-sensitive emission lines for direct-method abundances.
The larger scatter found for our KISS FMR compared to other studies may be due to the high-activity levels of galaxies found in KISS.
Deviations from the FMR seen at high redshift, where selection effects limit analysis exclusively to the most high-activity systems, may be a consequence of the same astrophysical mechanisms.
The increased scatter shown by the KISS galaxies may call into question the ``fundamental" nature of the FMR.
At least some galaxies deviate from the tight relation, a result reminiscent of that found by \citet{bib:Salim2014}.
\\
\indent With a self-consistent metallicity calibration and SED stellar mass estimates for all SFGs of the statistically representative KISS database, we gain the ability to pursue several interesting new science applications.
One possible project is a study comparing the star formation and abundance properties of galaxies as a function of environment.
Specifically, we plan to explore the properties of galaxies inside and outside of the Bo\"{o}tes Void, which is sampled by one of the KISS lists \citep{bib:Gronwall2004b}.
Our new O3N2-method abundance scale provides the ability to estimate metallicities for all KISS systems in this region.
The goal of this study would be to understand the root cause(s) of any observed differences in metallicity and/or star formation rate between systems within low- and high-density regions.
In a second application, we intend to create, for the first time, a so-called ``metallicity function" (MF).
The MF will examine the volume density of star-forming systems as a function of their metal abundance.
Only with a statistically representative sample of SFGs such as KISS can such a study be carried out.
Based on the classic Schechter luminosity function \citep{bib:Schechter1976}, we expect a plethora of low-luminosity systems to exist \citep{bib:Mateo1998}.
The \LZ\ relation then stipulates that these low-luminosity galaxies should be metal-poor.
Such low-abundance systems, however, are observationally uncommon (\citealp{bib:KunthSargent1983, bib:Izotov1997, bib:Hirschauer2016}).
An investigation of the MF of ELGs may shed light on this apparent paradox.
\\
\acknowledgements
\indent The early phases of the KISS project were supported by NSF grants AST 95-95020 and AST 00-71114 to JJS.
The current project received financial support from the Office of the Vice President for Research at Indiana University as well and the Indiana Space Grant Consortium, for which we are grateful.
Many colleagues from the KISS Team contributed to the data acquisition and analysis that made this project possible.
Special thanks to Caryl Gronwall, Anna Williams, Jessica Werk, Laura Chomiuk, Joanna Taylor, Duane Lee, Janice Lee, Jason Melbourne, and Matt Johnson.
We thank the anonymous referee for many helpful suggestions.
Finally, we wish to thank the support staffs at the observatories where the newer KISS spectroscopic data were obtained:
Kitt Peak National Observatory, Lick Observatory, the Hobby-Eberly Telescope, and MDM Observatory.
Without these dedicated individuals so much of what we do would not be possible.


\appendix

\section{A.\ Multiwavelength Photometric Data} 

\indent This Appendix describes the sources of photometric data and the processes by which we combined multiple datasets and verified the quality of all input data.

\subsection{A.1.\ Optical} 

\indent The KISS observations include flux measurements in the $B$ and $V$ filters for all sources (e.g., \citealp{bib:Salzer2000, bib:Salzer2001}).
Optical photometry in $ugriz$ filters from the Sloan Digital Sky Survey DR12 \citep{bib:Alam2015} is obtained using SQL queries to the CasJobs interface\footnote{\url{http://skyserver.sdss.org/CasJobs/}}.
We search for matching galaxies (\texttt{type==3}) within 3$''$ of all KISS objects, and use the \textsc{modelMag}s provided.
Approximately 98\% of KISS objects are matched in DR12.
\\
\indent For some galaxies with large angular sizes the SDSS deblending pipeline shreds them into multiple sources and poorly subtracts their sky levels \citep{bib:Blanton2011}.
The NASA Sloan Atlas (NSA\footnote{\url{http://www.nsatlas.org/}}) improves on the sky-subtraction and photometric measurements for nearby galaxies with large angular extents.
NSA also includes fluxes measured from \textit{GALEX} UV images using a self-consistent methodology.
We thoroughly compared the optical photometry from the SDSS catalog, NSA, and KISS $B$ and $V$ filters to determine the best optical fluxes for each galaxy.
\\
\indent As metrics of photometric integrity, we use both the positional offsets between SDSS and NSA measurements (at a threshold of $>$2$''$), and the flux differences between of KISS $BV$ photometry with estimated $BV$ fluxes from the SDSS $ugriz$ photometry (threshold of $>0.25$~mag, based on conversions from \citealp{bib:Jester2005}).
Additionally we perform SED fits (see \S2.3.2) including all available optical data and identify cases where the optical measurements from SDSS or NSA or KISS are inconsistent with the UV+IR SED.
\\
\indent For $\sim$60\% of targets, no NSA photometry is available and the SDSS fluxes are consistent with the KISS $BV$ photometry (and with the UV+IR SED fit).
For $\sim$30\% of targets we replace SDSS catalog photometry with superior NSA photometry and retain the KISS $BV$ fluxes.
We remove 21 KISS objects which are hopelessly blended with another source, making their photometric measurement intrinsically unreliable.
In the remaining cases, we keep either KISS, NSA, or SDSS optical photometry based on their agreement with preliminary SED fits.

\subsection{A.2.\ Ultraviolet} 

\indent Our UV data come from the \textit{GALEX} GR6\footnote{\url{http://galex.stsci.edu/GR6/}} and include Far UV (FUV, 1344-1786\AA) and Near UV (NUV, 1771-2831\AA) fluxes, with FWHM PSF resolutions of $4.3''$ and $5.3''$, respectively.
We use an SQL CasJobs query to find matches for $\sim$90\% of our targets within $6''$.
In a similar way to the optical comparisons, we use preliminary SED fits to test whether the GR6 UV fluxes or the NSA UV fluxes give lower reduced $\chi^2$ values on the SED fits, and also compare the S/N of GR6 and NSA fluxes.
For $\sim$7\% of targets, the GR6 UV photometry has better quality and produces better-fitting SEDs than the NSA UV measurements, so we adopt the GR6 values.

\subsection{A.3.\ Near Infrared} 

\indent Our NIR data come from 2MASS which observed the entire sky in $J$ (1.24$\mu$m), $H$ (1.66$\mu$m), and $K_S$ (2.16$\mu$m) NIR filters with $2''$ pixels \citep{bib:Skrutskie2006}.
We query \emph{both} the Point Source Catalog (PSC) and Extended Source Catalog (XSC) using a cone search within $10''$ of all KISS target positions.
This process resulted in the detection of 65\% of the KISS galaxies in the PSC in at least one band, while 40\% were found in the XSC.
At 2MASS resolution, some KISS galaxies are resolved while most are unresolved.
Merging these two catalogs gives the most complete NIR data available.
We use the standard-aperture corrected fluxes from the PSC (``stdap'') and the fully extrapolated XSC fluxes (``\_ext'').
For objects with faint apparent magnitudes (i.e., $J$ = 16~mag) the PSC and XSC flux measurements are equivalent, while for brighter (and larger) objects the PSC fluxes require a small additional correction in order to agree with the XSC fluxes.
Our final 2MASS photometry is designed to be as consistent and complete as possible, and includes NIR fluxes for $\sim$60\% of KISS.

\subsection{A.4.\ Mid Infrared} 

\indent Our MIR data come from WISE, which mapped the entire sky in w1 (3.4$\mu$m), w2 (4.6$\mu$m), w3 (12$\mu$m), and w4 (22$\mu$m) bands, with FWHM PSF resolutions of $6.1''$, $6.8''$, $7.4''$, and $12.0''$ respectively.
We matched our sources to the ALLWISE\footnote{\url{http://wise2.ipac.caltech.edu/docs/release/allwise/}} catalog.
This catalog only contains PSF profile-fit photometry, however, and some of the KISS targets are near enough to be resolved, making PSF photometry inaccurate. 
To mitigate this, we download ALLWISE Atlas images in all four bands for each target and use SExtractor \citep{bib:BertinArnouts1996} to measure their fluxes, and
also calculate distances to nearby neighbors which may affect the photometry (see \S3 in \citealp{bib:Janowiecki2017b} for a detailed description of this method and our treatment of blended sources and confusion).

\subsection{A.5.\ Summary of Photometric Data} 

\indent In the end, we construct SEDs using the optimal combination of all available photometry.
Table~\ref{tab:phot} summarizes the inventory of reliable flux measurements in each filter.
\\
\begin{deluxetable}{ccc} 
\tablewidth{0pt}
\tablecaption{Photometry inventory for KISS}
\tablehead{
Filter name & N$_\textrm{good}$ & Percentage 
}
\startdata
FUV & 1991 &  90\% \\
NUV & 1894 &  86\% \\
u   & 2133 &  97\% \\
g   & 2129 &  96\% \\
r   & 2133 &  97\% \\
i   & 2133 &  97\% \\
z   & 2130 &  97\% \\
B   & 2058 &  93\% \\
V   & 2058 &  93\% \\
J   & 1422 &  64\% \\
H   & 1339 &  61\% \\
K$_\textrm{s}$ & 1185 &  54\% \\
w1  & 1912 &  87\% \\
w2  & 1956 &  89\% \\
w3  & 1897 &  86\% \\
w4  & 1253 &  57\%
\enddata
\tablecomments{
N$_\textrm{good}$ is the number of flux measurements in each filter which pass all quality control tests.
}
\label{tab:phot}
\end{deluxetable}

\section{B.\ SED Fits and Systematic Verifications} 

\indent This Appendix first briefly describes the SED fitting process, and then the tests we performed to verify that SED fits with differing
wavelength coverage do not yield best-fit parameters which suffer from systematic differences.

\subsection{B.1.\ SED Fitting} 

\begin{figure} 
\center
\includegraphics[height=8.6cm,width=8.6cm]{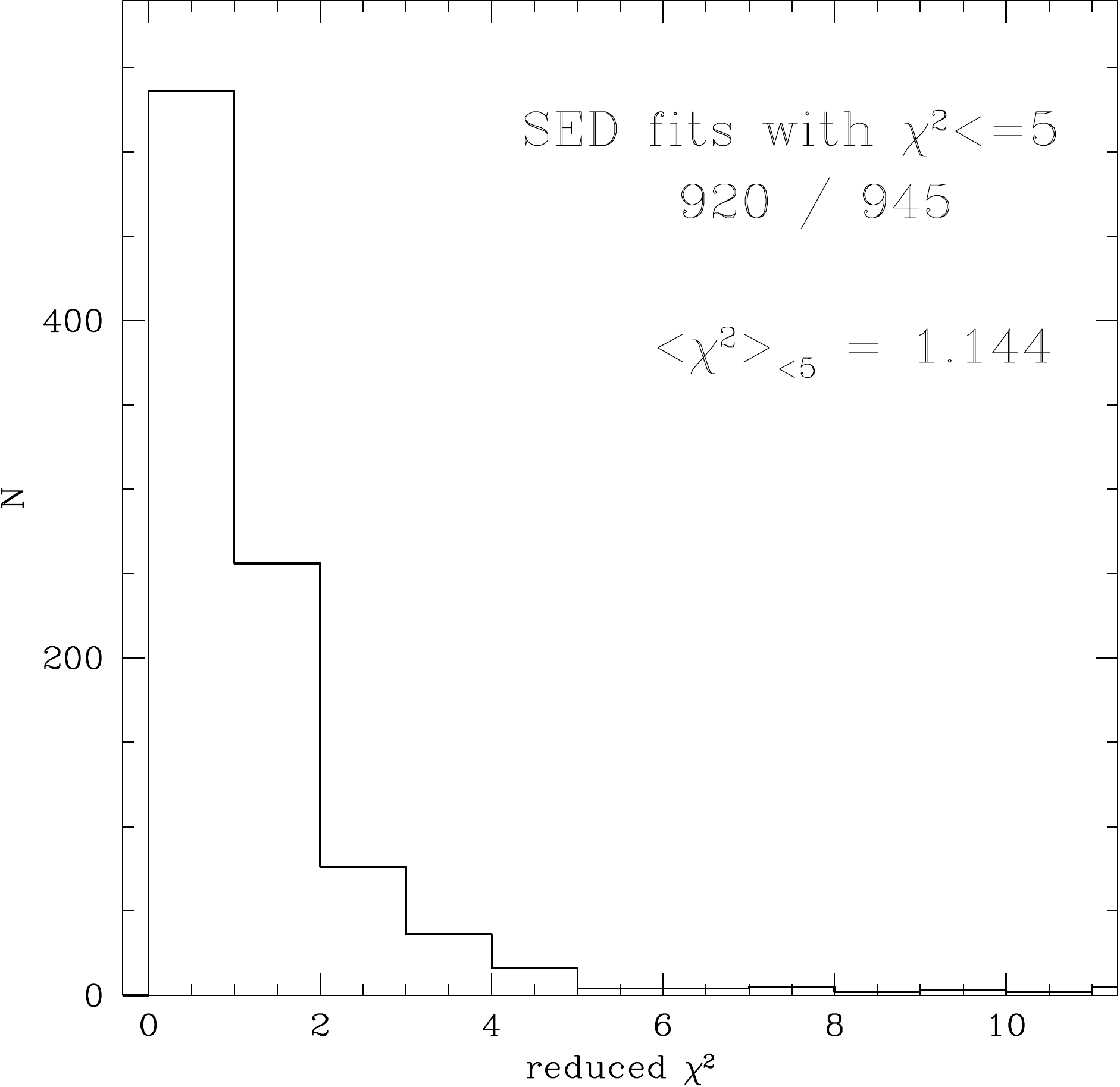}
\caption{Reduced $\chi^2$ distribution for KISS objects in ``Class~1'', with reliable flux measurements in all wavelength regimes.
More than $90$\% are good fits with $\chi^2<5$.
}
\label{fig:x2}
\end{figure}

\begin{figure} 
\center
\includegraphics[height=8.5cm]{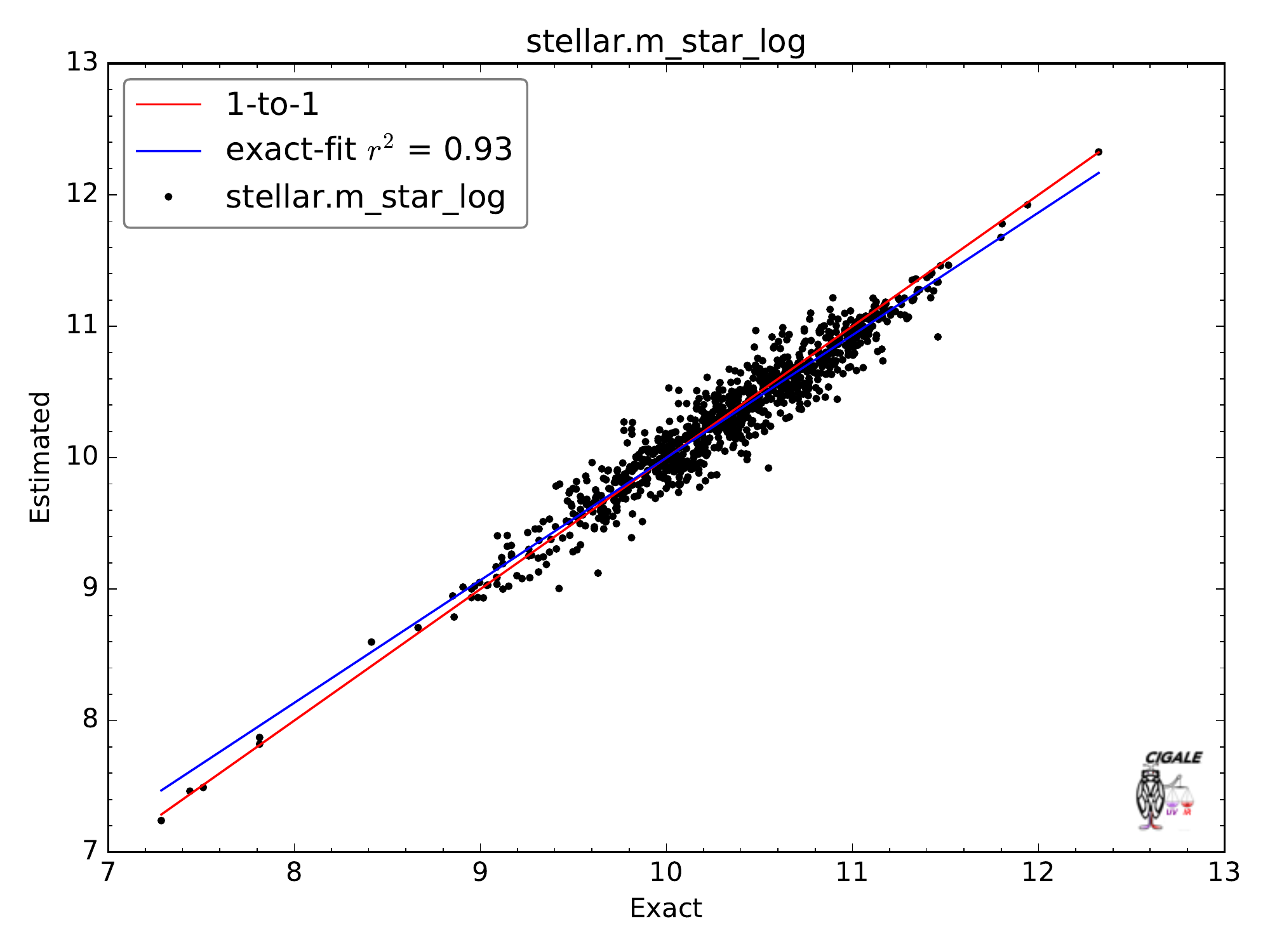}
\caption{Results of mock analysis of stellar mass for objects in ``Class~1'' SED fits.
}
\label{fig:mock}
\end{figure}

\indent We use CIGALE to fit SEDs following the same procedures as described in \S3 of \citet{bib:Janowiecki2017a}.
In short, CIGALE uses theoretical models to derive the flux emitted and absorbed by stars, gas, and dust in a grid of galaxy models.
All parameters used in this grid are given in Table~\ref{tab:grid}.
The stellar populations synthesis models \citep{bib:BruzualCharlot2003} use a \citet{bib:Salpeter1955} IMF and are sampled between metallicities of $Z=0.0001$ and $Z=0.05$.
We use two-burst stellar population tau-models where the star formation histories of the old and young populations are described by declining exponential functions with well-sampled scale times and ages, and are related to each other by a mass ratio.
\\
\begin{deluxetable}{cccc} 
\tablecaption{SED-fitting grid parameters}
\tablewidth{0pt}
\tablehead{ Parameter & Symbol & Values & Units }
\startdata
Old pop. age [Myr]              & age$_o$    &  2000, 7000, 13000  & Myr\\
Old pop. e-folding time [Myr]   & $\tau_o$   &  100, 1000, 10000 & Myr\\
Young pop. age [Myr]            & age$_y$    &  20, 50, 150, 500, 2000 & Myr \\
Young pop. e-folding time [Myr] & $\tau_y$   &  3, 30, 300, 3000 & Myr \\
Young pop. mass fraction        & $f_b$      &  0.001, 0.003, 0.01, 0.1, 0.5, 0.99 & \nodata \\
Stellar metallicity             & $Z$        &  0.0004, 0.004, 0.008, 0.02, 0.05 & $Z/Z_\odot$ \\
Amount of dust attenuation      & $E(B-V)_y$ &  0.01, 0.03, 0.065, 0.1, 0.25, 0.4, 0.7 & \nodata \\
Power-law slope on extinction law & $\delta$ &  −0.7, −0.3, −0.1, 0, 0.1, 0.3, 0.7 & \nodata \\
Dust heating parameter          & $\alpha$   &  1, 1.5, 2, 2.5 & \nodata
\enddata
\tablecomments{Note that we require that age$_o$ $>$ age$_y$ + 10 Myr, to enforce a separation between the old and young stellar populations. 
}
\label{tab:grid}
\end{deluxetable}
\indent Metallicity-dependent nebular emission and absorption is computed following \citet{bib:Inoue2011}, and dust attenuation is based on the method of \citet{bib:Cardelli1989}, with the formulas from \citet{bib:Calzetti2000} and \citet{bib:Leitherer2002}.
Dust (re-)emission is modeled by templates from \citet{bib:Dale2014}, which depend on a single heating parameter, which we sample in our grid.
We have not included any AGN contributions.
\\
\indent After generating a grid of synthetic SEDs, CIGALE compares each model to the observed SED and computes a $\chi^2$ value at each grid point.
Figure~\ref{fig:x2} shows the reduced $\chi^2$ distribution for objects in ``Class~1'' (see \S\ref{sec:class} for more details).
These $\chi^2$ values are used to generate probability distribution functions (PDFs) for each analyzed parameter, in a Bayesian-like framework \citep{bib:Kauffmann2003, bib:Salim2005, bib:Salim2007, bib:Noll2009}.
For further discussion of degeneracies, dependencies, and reliability estimates of various parameters in SED fits, see \S3 of \citet{bib:Janowiecki2017a}.
\\
\indent We briefly discuss one of the methods used to assess and measure the reliability and robustness of the fits.
Figure~\ref{fig:mock} shows the results of our mock fitting tests (see \citealp{bib:Janowiecki2017a} for more details).
We use each synthetic SED from our grid (with known parameters) as ``mock observations'' which we fit and derive best-fit parameters.
The best-fit parameters from the ``mock fits'' are compared with the true input parameter values to see how reliably certain parameters are recovered.
While there is significant scatter in the total stellar masses (0.12 dex rms), they are reliably recovered in the mock fits. 
\\

\subsection{B.2.\ Systematic Verifications} 
\label{sec:class}

\begin{deluxetable}{ccc} 
\tablewidth{0pt}
\tablecaption{Classes of SED fits}
\tablehead{ Class & N$_\textrm{gal}$ & Filters }
\startdata
1 & 913 & UV+BV+ugriz+JHK+W123+W4 \\
2 & 315 & UV+BV+ugriz+JHK+W123\textcolor{white}{+W4} \\
3 & 587 & UV+BV+ugriz\textcolor{white}{+JHK}+W123\textcolor{white}{+W4} \\
4 & 35  & UV+BV\textcolor{white}{+ugriz+JHK}+W123\textcolor{white}{+W4} \\
5 & 131 & \textcolor{white}{UV+}BV+ugriz\textcolor{white}{+JHK}+W123\textcolor{white}{+W4} \\
6 & 116 & \textcolor{white}{UV+BV+}ugriz\textcolor{white}{+JHK}+W123\textcolor{white}{+W4} \\
7 & 67  & \textcolor{white}{UV+BV+}ugriz\textcolor{white}{+JHK+W123+W4} \\
0 & 43  & unclassified (21 are blends)
\enddata
\tablecomments{\\
\small \raggedright
UV - UV fluxes from \textit{GALEX} GR6/NSA,\\
BV - $B$ and $V$ fluxes from KISS database,\\
ugriz - $ugriz$ fluxes from SDSS DR12/NSA\\
JHK - $JHK_s$ fluxes from 2MASS PSC/XSC\\
W123 - \textit{WISE} w1,w2,w3 fluxes\\
W4 - \textit{WISE} w4 flux\\
}
\label{tab:class}
\end{deluxetable}

\begin{figure} 
\center
\includegraphics[height=8.6cm,width=8.6cm]{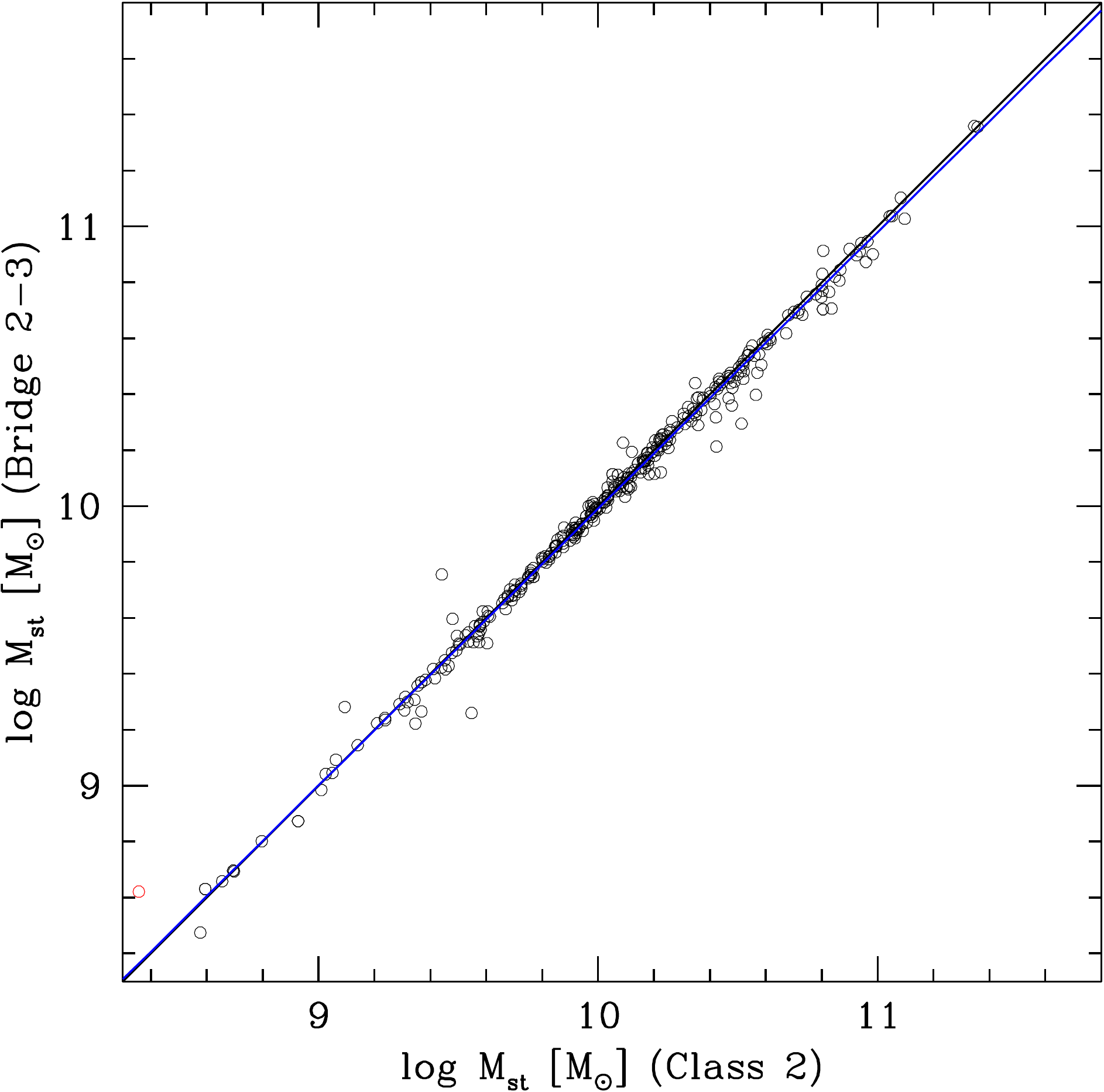}
\caption{Comparison between stellar mass from Class~2 (x-axis) and Bridge~2-3 (y-axis).
We show the least squares bisector fit (in blue), which is nearly indistinguishable from the black unity line.
}
\label{fig:b23}
\end{figure}

\indent As described in Appendix~A, our photometric dataset is heterogeneous.
Some galaxies have flux measurements at 16$+$ wavelengths, while others only have a few data points.
In order to verify that different combinations of observations do not bias our determinations of stellar mass across the KISS sample, we embark on a series of comparisons and tests to explore possible systematic offsets.
Toward that end, we devise a series of ``classes'' of SEDs, where each class has a different combination of photometric data available.
Each KISS galaxy is assigned to one class.
The classes (and number of galaxies in each) are given in Table~\ref{tab:class}.
\\
\indent Encouragingly, ``Class~1'' includes galaxies with flux measurements in all wavelength regimes, and comprises $\sim$40\% of the KISS sample.
This first class represents the best-possible SEDs available for this comparison, and should produce the most well-constrained estimates of stellar mass (and was used to show sample results in previous figures).
Subsequent classes include objects with progressively fewer flux measurements, but their SED fits can still provide robust stellar mass estimates.
Within each class, we fit SEDs to the KISS objects using only the filters of that class.
\\
\indent To compare results between classes, we use a set of ``Bridge'' samples.
For example, ``Bridge~1-2'' is used to connect the stellar masses from the first and second class fits.
This bridge sample includes all of the objects from Class~1, but fits only only the filters included in Class~2.
That allows us to compare the fits of Class~1 objects (using the complete set of filters) with fits to the same objects using the reduced number of filters in Class~2, to quantify any systematic differences between the fits.
\\
\indent As an example, if removing 2MASS JHK photometry were to introduce a systematic difference in stellar mass estimates, then the Class~2 objects (including JHK fluxes) will have different masses when they are fit in Bridge~2-3, using Class~3 filters (not including JHK fluxes).
In this example, the same 315 galaxies are fit with UV+BV+ugriz+JHK+W123 data (Class~2 filters) and again without JHK (Bridge~2-3:\ Class~2 objects, using Class~3 filters).
We then check for any differences in the best-fit results for stellar mass.
Included in Figure~\ref{fig:b23} is the mass comparison for Bridge~2-3.
\\
\indent While there is a scatter of $\sim$3\%, removing the JHK fluxes introduces no significant systematic effects on the determination of stellar mass.
Nonetheless, we use the bivariate least squares fit to generate a corrected mass to bring Class~3 objects onto the same scale as Class~2 fits, as follows:
\[
\text{log}~\text{M}_{*}(\text{Bridge}~2-3) = 0.993 \times \text{log}~\text{M}_{*} (\text{Class}~2) + 0.083
\]
\indent We use this system of Bridges and Classes to sequentially link all of the stellar mass estimates together from each set of flux combinations, and to remove any systematic offsets (relative to Class~1 determinations).
Note that Class~4 does not follow the pattern of dropping one wavelength regime in each subsequent class.
As such, it is not used in the chain of bridges between classes.
Instead, Class~4 is included to prevent the loss of these 35 galaxies which have a significant number of flux measurements, but lack good SDSS $ugriz$ fluxes.
If not for their inclusion in Class~4, these 35 galaxies would drop to Class~0, which means the SED fits use any available fluxes and no effort is made to calibrate potential systematic offsets in stellar mass.
\\
\indent In all, we use the Bridges to link each Class of SED fits and homogenize all of our stellar masses.
Given the robustness of SED fits even with limited flux measurements, however, these mass corrections are small.
Most of the corrections applied are $\lesssim0.5\%$, and the largest are $<2\%$.
\\
\\
\\
\\
\\
\\


\end{document}